\documentclass[twocol]{ametsocV5}

\usepackage{amsmath,amsfonts,amssymb,bm}
\usepackage{mathptmx}
\usepackage{newtxtext}
\usepackage{newtxmath}
\usepackage{cuted}
\newcommand{\pder}[2][]{\frac{\partial#1}{\partial#2}}
\newcommand{\Dder}[2][]{\frac{D#1}{D#2}}
\newcommand{\dpder}[2][]{\frac{\partial^2#1}{\partial#2^2}}

\title{On the Effect of Surface Friction and Upward Radiation of Energy on Equatorial Waves \\
\textbf{\color{red} Submitted for peer review.}}

\authors{Jonathan Lin\correspondingauthor{Jonathan Lin, jzlin@mit.edu}}

\affiliation{Lorenz Center, Department of Earth, Atmospheric, and Planetary Sciences,\\Massachusetts Institute of Technology, Cambridge, Massachusetts}

\extraauthor{Kerry Emanuel}
\extraaffil{Lorenz Center, Department of Earth, Atmospheric, and Planetary Sciences,\\Massachusetts Institute of Technology, Cambridge, Massachusetts}

\abstract{In theoretical models of tropical dynamics, the effects of both surface friction and upward wave radiation through interaction with the stratosphere are oft-ignored, as they greatly complicate mathematical analysis. In this study, we relax the rigid-lid assumption and impose surface drag, which allows the barotropic mode to be excited in equatorial waves. In particular, a previously developed set of linear, strict quasi-equilibrium tropospheric equations is coupled with a dry, passive stratosphere, and surface drag is added to the troposphere momentum equations. Theoretical and numerical model analysis is performed on the model in the limits of an inviscid surface coupled to a stratosphere, as well as a frictional surface under a rigid-lid. This study confirms previous research that shows the presence of a stratosphere strongly shifts the growth rates of fast propagating equatorial waves to larger scales, reddening the equatorial power spectrum. The growth rates of modes that are slowly propagating and highly interactive with cloud-radiation are shown to be negligibly affected by the presence of a stratosphere. Surface friction in this model framework acts as purely a damping mechanism and increases the poleward extent of the equatorial waves through barotropic vorticity generation. Numerical solutions of the coupled troposphere-stratosphere model with surface friction also show that the barotropic mode can be tropospherically trapped when excited by surface friction but in the presence of a highly stratified stratosphere. The superposition of phase-shifted barotropic and first baroclinic modes is also shown to lead to an eastward vertical tilt in the dynamical fields of Kelvin-wave like modes.}

\begin{document}
 
\maketitle

%
%
%

%








\section{Introduction}

Reduced models of the tropical atmosphere have found much success in replicating many characteristics of the tropical circulation. Of prominence are theoretical models that reduce the linear, primitive equations to the shallow water equations by use of only the first baroclinic mode~\citep{matsuno1966quasi, gill1980some}. Indeed, remarkable evidence of the linear and neutral equatorial waves that arise from the Matsuno-Gill model have been documented in the tropics~\citep{wheeler1999convectively}.  The first baroclinic mode has also been used extensively in simple models of the tropics, from studies of intraseasonal oscillations \citep{emanuel1987air, sobel2001weak} to steady circulations~\citep{neelin1987modeling, emanuel1994large, neelin2000quasi, sobel2000modeling}, among many others. 

While the first baroclinic mode is certainly a prominent feature of the tropical atmosphere \citep{xu1989tropical}, observational data and analysis have suggested the existence of another vertical mode, the second (stratiform) baroclinic mode~\citep{mapes1995diabatic, straub2002observations}. These observations have lead to a plethora of theoretical studies that analyze how the interaction between the first and second baroclinic modes can lead to instability in equatorial waves~\citep{mapes2000convective, kuang2008moisture}.

Perhaps curiously left behind is the barotropic mode, even though it does indeed survive the strict quasi-equilibrium assumption; in fact strict quasi-equilibrium eliminates all but the barotropic and first baroclinic modes~\citep{emanuel1987air, neelin2000quasi}. The barotropic mode can be excited in linear models of equatorial waves through coupling to the first baroclinic mode via surface friction \citep{wang1988dynamics, wang1990dynamics, moskowitz2000analysis} or removal of the rigid-lid assumption~\citep{yano1991improved}. In this study, we build on the linear, strict quasi-equilibrium model first formulated in \citet{khairoutdinov2018intraseasonal} (henceforth, KE18) and further analyzed in \citet{emanuel2020slow} (henceforth, E20), and investigate how excitation of the barotropic mode through both surface friction and coupling to the stratosphere affects the growth, structure, and propagation of equatorial waves.

Baked into the modal decomposition of many simple models of the tropics is the rigid lid assumption, since it dramatically simplifies analytic solutions. However, in reality the tropopause does not act as a rigid lid on the troposphere. While it is true the stratosphere has a larger stratification than the troposphere, the stratification in the stratosphere is not infinite. A ``leaky lid" analogy is more accurate, as wave energy can radiate to the stratosphere.

A few studies have investigated the impact of a stratification jump at the tropopause on the evolution of a wave in the atmosphere. \citet{yano1991improved} extends the tropical intraseasonal model introduced in \citet{emanuel1987air} by imposing a dry, passive stratosphere above the troposphere, and found that adding a stratosphere strongly damps the smallest scale $v = 0$ waves, shifting the growth rates to the larger scales. By imposing wave-radiation boundary conditions, other studies have found similar effects: that the effect of the stratosphere is a dampening one~\citep{moskowitz2000analysis, kuang2008modeling}. \citet{chumakova2013leaky} further investigates the leaky lid effect by deriving a set of vertical, dissipative modes using the 2-D, linear, Boussinesq equations, overlaying a stratosphere with buoyancy frequency $N_2$ over a troposphere with buoyancy frequency $N_1$, where $N_2 > N_1$. In their model, a new barotropic-like mode appears with a fast damping time scale. However, the vertical modes of \citet{chumakova2013leaky} are not orthogonal; they have also been criticized as unphysical, since the energy is unbounded with height, and they also do not admit steady state solutions to steady state heating~\citep{edman2017beyond}. Regardless of the exact specifics in applying a leaky-lid above a troposphere, the inclusion of a stratosphere tends to shift growth rates of unstable modes to larger scales and allows the barotropic mode to be excited.

There have also been many studies that have investigated the role surface friction plays in modifying equatorial waves, and more prominently, the Madden-Julian Oscillation (MJO). These theories based on CISK (conditional instability of the second kind), in which surface friction was postulated to act, through its induced moisture convergence, as a destabilizing mechanism for convectively coupled equatorial waves~\citep{wang1988dynamics, wang1990dynamics, moskowitz2000analysis}. \citet{wang1988dynamics} and \citet{wang1990dynamics} formulated a 2.5 layer ``frictional WAVE-CISK" model (2 tropospheric layers and a thin frictional boundary layer) in which the barotropic mode can be excited through surface friction. The surface friction acts to induce vertical motion at the top of the boundary layer, which can amplify wave-disturbances if correlated with temperature anomalies. CISK theories have received much criticism [see \citet{emanuel1994large} and \citet{neelin1994modes}], since they violate the convective statistical equilibrium hypothesis, where the rate of production of convectively available potential energy (CAPE) by the large-scale environment is very nearly balanced by its consumption via convection~\citep{arakawa1974interaction}. In the context of surface friction, this means that surface convergence is a by-product and not a driver of convection. Indeed, observations in the tropics support the statistical equilibrium model for convection in the tropics~\citep{betts1982saturation, xu1989tropical}. Further, numerical simulations of large-scale equatorial waves do not support the idea that surface friction acts as a destabilizing mechanism for large-scale equatorial waves~\citep{chao2001role}.

Surface friction and boundary layer convergence have also been cited as one mechanism for moistening of the lower troposphere by shallow upward motion east of the MJO center~\citep{wang1988dynamics, hsu2012role, adames2014three}. In theoretical models, this mechanism has been shown to influence the propagation speed and growth rates of the MJO through modulation of the gross moist stability~\citep{sobel2013moisture, adames2016mjo}.

In this study, we will show how surface friction and interaction with the stratosphere through upward wave energy radiation modifies the characteristics of equatorial waves. The rigid-lid assumption is removed by explicitly coupling a passive and dry stratosphere on top of a convecting troposphere, and the energy density of solutions is enforced to decay with height. Surface drag is imposed on a thin boundary layer at the surface. In particular, we will focus on how the two aforementioned mechanisms can excite the barotropic mode. More uniquely, the barotropic mode in this study does not separate convective heating from large-scale thermodynamics, and thus it does not violate the strict quasi-equilibrium hypothesis. 

The paper is organized as follows. Section~\ref{sec_linear_model} describes the linear model. Section~\ref{sec_solutions} presents the solutions of the linear model. The paper concludes with a discussion and summary in section~\ref{sec_summary}.

\section{Linear model \label{sec_linear_model}}
In this section, we describe and formulate the governing equations of our linear model. The model formulation is separated into two parts, section~\ref{sec_linear_model}\ref{sec_quasi} which describes the tropospheric model, and section~\ref{sec_linear_model}\ref{sec_coupling} which describes the stratospheric model.

\subsection{Strict quasi-equilibrium troposphere \label{sec_quasi}}
KE18 and E20 derive and analyze an equation set for a linear system that describes the dynamics and thermodynamics of an atmosphere that maintains a vertically constant saturation moist entropy $s^*$ in the free troposphere. However, in both of those studies, the authors assume a rigid-lid and frictionless surface. Hence, in their model, only the baroclinic mode can be excited, and upward radiation into the stratosphere is absent. 

Here we derive nearly equivalent dynamics, but remove the rigid-lid hypothesis and include the barotropic mode. To begin, we first apply a Galerkin decomposition of the vertical modes of the troposphere and truncate all modes except the first two basis functions, $V_0$ and $V_1$, which are defined as the barotropic and baroclinic modes, respectively~\citep{neelin2000quasi}. Mathematically, they are:
\begin{align}
    V_0 &= 1 \\
    V_1 &= \frac{\overline{T}(p) - [\overline{T}]}{T_{b} - [\overline{T}]}
\end{align}
where $\overline{T}$ is the basic state temperature, $T_b$ is the boundary layer temperature, $[\overline{T}]$ is the pressure-weighted vertical average of temperature. The operator $[ \: \bullet \: ] = \frac{1}{\Delta p} \int_{p_t}^{p_s} \bullet \: \: dp$ is the pressure-weighted vertical average along a moist adiabat, where $p_s$ is the surface pressure, $p_t$ is the tropopause pressure, and ${\Delta p = p_s - p_t}$. As is standard for vertical modes, the basis functions are orthogonal, or $\int_{p_s}^{p_t} V_0 V_1 dp = 0$. Furthermore, note that $[V_1] = 0$. From this vertical mode decomposition, we assume separable dependencies between the horizontal and vertical modes as follows:
\begin{align*}
    \phi(x, y, p, t) &= \phi_0 (x, y, t) V_0 + \phi_1 (x, y, t) V_1(p)
\end{align*}
and likewise for the other prognostic variables.

For a strict quasi-equilibrium troposphere in which the saturation moist entropy $s^*$ is constant with height, linearized geopotential perturbations are directly connected to $s^*$ perturbations~\citep{emanuel1987air}.
\begin{equation}
    \pder[\phi^\prime]{p} = -\bigg( \pder[T]{p} \bigg)_{s^*} s^{*\prime} \label{eq_dphi_dp}
\end{equation}
where prime superscripts indicate perturbation quantities. The above may be directly integrated from the surface upwards to yield:
\begin{equation}
    \phi^\prime(p) = \phi^\prime_b + s^{*\prime} (\overline{T}_b - \overline{T}(p)) \label{eq_phi_qe}
\end{equation}
where $\phi^\prime_b$ is the geopotential in the boundary layer. When non-dimensionalized (see Appendix A for details), Equation (\ref{eq_phi_qe}) yields:
\begin{equation}
    \phi^\prime(p) = \phi_b^\prime + (1 - V_1) s^{*\prime} \label{eq_phi}
\end{equation}
Note, the geopotential can be separated into its barotropic and baroclinic components:
\begin{align}
    \phi_0(x, y, t) &= (\phi_{b}^\prime + s^{*\prime}) V_0 \\
    \phi_1(x, y, p, t) &= - s^{*\prime} \, V_1
\end{align} The pressure-weighted vertical average operator is also applied to Equation (\ref{eq_phi_qe}) to give:
\begin{equation}
    \phi^\prime_b = [\phi]^\prime + s^{*\prime} ([\overline{T}] - \overline{T}_b) \label{eq_phib}
\end{equation}
which in non-dimensional form is:
\begin{equation}
    \phi^\prime_b = \phi^\prime_0 - s^{*\prime}  \label{eq_phib_phi0}
\end{equation}
Unlike the purely baroclinic motions described in \cite{khairoutdinov2018intraseasonal}, the geopotential now contains contributions from the barotropic mode. Note that Equation (\ref{eq_phi}) can be evaluated at the tropopause and combined with Equation (\ref{eq_phib_phi0}) to obtain the tropopause geopotential $\phi^\prime_{\text{tp}}$, which will be required to couple the system to the stratosphere:
\begin{equation}
    \phi^\prime_{\text{tp}} = \phi^\prime_0 - V_1(p_t) s^{*\prime} \label{eq_phi_tp}
\end{equation}
where $p_t$ is the non-dimensional tropopause pressure.

Next, we formulate the full equation of motion on an equatorial $\beta$-plane, adding in surface friction, which is represented as applying drag on an infinitesimally small boundary layer.
\begin{align}
    \Dder[\boldsymbol{V}]{t} &= -\nabla \phi - \hat{k} \times \beta y \boldsymbol{V} - \delta(p - p_s) \frac{C_d}{h_b} |\boldsymbol{V}| \boldsymbol{V} 
\end{align}
where $\boldsymbol{V}$ is the vector wind, $\beta$ is the meridional gradient of the Coriolis force, $\delta$ is the Dirac delta function, $p$ is pressure, $C_d$ is the drag coefficient, and $h_b$ is the boundary layer depth. Note the surface stress is parameterized using the bulk aerodynamic drag formula. Linearizing around surface easterlies, non-dimensionalizing according to details in Appendix A, substituting in Equation (\ref{eq_phi}), and dropping all primes of perturbation quantities, we obtain:
\begin{align}
    \pder[u]{t} &= -\pder[\phi_b]{x} + (1 - V_1) \pder[s^*]{x} + y v - 2 F u \delta(p - p_s) \\
    \frac{1}{\delta_x} \pder[v]{t} &= -\pder[\phi_b]{y} + (1 - V_1) \pder[s^*]{y} - y u - \frac{F}{\delta_x} v \delta(p - p_s) 
\end{align}
where $F$ is the non-dimensional surface friction coefficient and $\delta_x$ represents the magnitude of zonal geostrophy [corresponding to $\delta$ in KE18]. Finally, we project the linearized horizontal momentum equations onto the barotropic and baroclinic modes:
\begin{align}
    \pder[u_0]{t} &= -\pder[\phi_0]{x} + y v_0 - 2 F (u_0 + u_1) \label{eq_uBT} \\
    \frac{1}{\delta_x} \pder[v_0]{t} &= -\pder[\phi_0]{y} - y u_0 - \frac{F}{\delta_x} (v_0 + v_1) \label{eq_vBT} \\
    \pder[u_1]{t} &= \pder[s^*]{x} + y v_1 - 2 F (u_0 + u_1) \label{eq_uBC} \\
    \frac{1}{\delta_x} \pder[v_1]{t} &= \pder[s^*]{y} - y u_1 - \frac{F}{\delta_x} (v_0 + v_1) \label{eq_vBC}
\end{align}
Next, we enforce mass continuity through the continuity equation in pressure coordinates:
\begin{equation}
    \pder[u]{x} + \pder[v]{y} + \pder[\omega]{p} = 0
\end{equation}
where $\omega$ is the pressure vertical velocity. Integrating the continuity equation from the surface to the tropopause and using the fact that $[V_1] = 0$:
\begin{equation}
    \int_{p_s}^{p_t} \nabla_H \cdot \boldsymbol{V} = \pder[u_0]{x} + \pder[v_0]{y} = \omega(p_t) \label{eq_omega_cont}
\end{equation}
where we have used zero vertical velocity condition at the lower boundary, and $\omega(p_t)$ is the tropopause vertical velocity. Equation (\ref{eq_omega_cont}) shows that the tropopause vertical velocity is only a function of the divergence of the barotropic mode, as the baroclinic mode is zero at the tropopause, by definition.

Finally, the thermodynamic equations in the troposphere link the dynamics to the thermodynamics, and are only slightly modified from KE18 and E20 in that horizontal diffusion is removed and $\kappa = 1$:\footnote{The second author discovered that $\kappa$, an additional non-dimensional coefficient that scales the cloud-radiative feedback and was defined in KE18 and E20, must be equal to 1 for consistency of the non-dimensional scaling.}
\begin{align}
    \pder[s^*]{t} &= (1 + C) s_m - w - \alpha u_b - \chi s^* \label{eq_s} \\
    \gamma \pder[s_m]{t} &= -D s^* - \alpha u_b - G w + C s_m \label{eq_sm}
\end{align}
where $s_m$ is a characteristic moist entropy of the free troposphere, $w = -\pder[u_b]{x} - \pder[v_b]{y}$ is a proxy for the mid-level vertical velocity based on the boundary layer zonal velocity, $u_b = u_0 + u_1$, and boundary layer meridional velocity $v_b = v_0 + v_1$. The non-dimensional coefficients $C$, $\alpha$, $\chi$, $D$, $G$, and $\gamma$ are described and formulated in detail in KE18. Briefly, $C$ represents the strength of cloud radiative feedback, $\alpha$ is the wind-induced surface heat exchange (WISHE) feedback parameter, $\chi$ is boundary layer damping, $D$ is entropy damping, $G$ is the gross moist stability, and $\gamma$ modifies the time scale of tropospheric entropy.

Equations (\ref{eq_uBT}) - (\ref{eq_vBC}), (\ref{eq_omega_cont})- (\ref{eq_sm}) formulate the tropospheric system, where $u_0$, $v_0$, $u_1$, $v_1$, $\phi_0$, $s$, $s_m$ are the unknown variables. Note that the linear system is not complete: additional specification of the vertical velocity at the tropopause is required to complete the system. As mentioned previously, studies that assume a rigid-lid set the tropopause velocity to be zero. Other studies parameterize the tropopause dynamics using a wave-radiation upper boundary condition~\citep{moskowitz2000analysis, kuang2008modeling}. In this study, we take a different approach and couple the vertical velocity to an explicit stratosphere model, as will be derived in section~\ref{sec_linear_model}\ref{sec_coupling}.

\subsection{Coupling to the stratosphere \label{sec_coupling}}
In this section, we couple a dry, passive stratosphere to the strict quasi-equilibrium troposphere described in section~\ref{sec_linear_model}\ref{sec_quasi}. We choose to represent a dry and passive stratosphere using the linearized, inviscid primitive equations in log-pressure coordinates and in hydrostatic balance [see Chapter 3 of~\citet{andrews1987middle}]:
\begin{align}
    \pder[u^\prime_s]{t} = -\pder[\phi^\prime_s]{x} + \beta y v^\prime_s \label{eq_strat_U} \\
    \pder[v^\prime_s]{t} = -\pder[\phi^\prime_s]{y} - \beta y u^\prime_s \label{eq_strat_V} \\
    \pder[u^\prime_s]{x} + \pder[v^\prime_s]{y} + \frac{1}{\rho_s} \pder[(\rho_s w^{*\prime}_s)]{z^*} = 0 \label{eq_strat_cont} \\
    \pder[]{t} \pder[\phi^\prime_s]{z^*} + w^{*\prime}_s N^2 = 0 \label{eq_strat_T}
\end{align}
where subscripts of $s$ denote quantities in the stratosphere, $w^*_s$ is the log-pressure vertical velocity, $N^2$ is the buoyancy frequency, $\rho_s$ is the basic state density, and the log-pressure vertical coordinate ${z^* \equiv - H \ln ( p / p_t ) + 1}$ is defined such that $z^* = 1$ is the bottom boundary, or the tropopause. Equations (\ref{eq_strat_U})-(\ref{eq_strat_T}) are non-dimensionalized according to notation shown in Appendix A, with the additional specification that the non-dimensional density decays exponentially with a scale height $H_s$. The resulting, non-dimensional equations are shown in Equations (\ref{eq_strat_U_nd}) - (\ref{eq_strat_rho}), with primes removed from perturbation quantities.
\begin{align}
    \pder[u_s]{t} = -\pder[\phi_s]{x} + y v_s \label{eq_strat_U_nd} \\
    \frac{1}{\delta_x} \pder[v_s]{t} = -\pder[\phi_s]{y} - y u_s \label{eq_strat_V_nd} \\
    \pder[u_s]{x} + \pder[v_s]{y} + \frac{1}{\rho_s} \pder[(\rho_s w^*_s)]{z^*} = 0 \label{eq_strat_cont_nd} \\
    \pder[]{t} \pder[\phi_s]{z^*} + w^*_s S = 0 \label{eq_strat_T_nd} \\
    \rho_s = \exp \Big( \frac{H}{H_s}(1 - z^*) \Big) \label{eq_strat_rho}
\end{align}
where $S$ is a non-dimensional stratospheric stratification.

It is important to note that these equations form a complete system by themselves. The stratospheric linear system admits neutral equatorial wave solutions under a rigid-lid upper boundary condition. However, under a upward wave radiation boundary condition, all of the solutions decay exponentially in time since there is no forcing in the stratosphere model and wave energy escapes upwards. Growing solutions that satisfy the upward wave radiation boundary condition in the stratosphere are possible, however, if there is mechanical forcing from the troposphere via the tropopause. In order to investigate these kinds of solutions, we must couple the troposphere system with the stratosphere system in a consistent fashion. Classical coupling conditions require continuity of normal stress across the interface, and continuity of normal displacement to the fluid interface. Since the free-troposphere is modeled as inviscid, the first condition simplifies to continuity of pressure:
\begin{equation}
    \phi_s(x, y, z^* = 1, t) = \phi(x, y, p = p_t, t) \label{eq_phi_match}
\end{equation}
Since there is no imposed shear across the tropopause, the second condition implies continuity of vertical velocity:
\begin{equation}
    w^*_s(x, y, z^* = 1, t) = -B \omega(x, y, p = p_t, t) \label{eq_w_match}
\end{equation}
where $B = ((p_s - p_t)/ p_t)(H_s / H) > 0$ is a conversion coefficient between pressure coordinates and log-pressure coordinates.

\subsection{Full, linear model}
The troposphere system [Equations (\ref{eq_uBT}) - (\ref{eq_vBC}), (\ref{eq_omega_cont}) - (\ref{eq_sm})] is coupled to the stratosphere system [Equations (\ref{eq_strat_U_nd}) - (\ref{eq_strat_rho})] through the two matching conditions [Equations (\ref{eq_phi_match}) and (\ref{eq_w_match})]. Altogether, these formulate a complete linear system, in which growing solutions whose energy decays to zero as $z^* \rightarrow \infty$ represent modes that grow in the troposphere and propagate vertically into the stratosphere.

\section{Solutions \label{sec_solutions}}
The full linear model is a complex system that cannot be easily solved theoretically. However, analyzing the model in the limits of (1) an inviscid surface with coupling to the stratosphere, and (2) a frictional surface under a rigid-lid, allows us to isolate the impacts of both mechanisms. Solutions of the full model, with active surface friction and stratosphere coupling, are then analyzed to illuminate their combined effects.

\subsection{Leaky modes \label{sec_leaky_theory}}
In this section, we focus first on solutions of the purely leaky modes, with no surface friction ($F = 0$). The solutions are analyzed separately: modes where $v = 0$ and higher order meridional modes where there is non-zero meridional velocity. In what follows, unless otherwise stated, the primes are dropped from the linear perturbation variables.

\subsubsection{$v = 0$ modes}
Although the full linear model is extremely complex, restricting the solutions to $v = 0$ allows for tractable analytical insight. In the troposphere, we assume solutions of the form:
\begin{equation}
    u_0 = \hat{U}_0 (y) \exp(ikx + \sigma t)
\end{equation}
where $k$ is the zonal wave number, $\sigma$ is the complex growth rate, and capitalized variables with hat notations are the meridional structure functions. Equivalent forms are assumed for $\phi_0$, $u_1$, $s$, $s_m$. Solutions in the stratosphere are assumed of the form:
\begin{equation}
    u_s = \frac{\hat{U_s}(y)}{\sqrt{\rho_0}} \exp(ikx + \sigma t + im(z^*-1)) \label{eq_strat_sol}
\end{equation}
where $m$ is the complex vertical wavenumber. Equivalent forms are also assumed for $\phi_s$, and $w^*_s$. As in KE18, the meridional structure $Y$ of the $v = 0$ modes in the troposphere can be derived by combining Equations (\ref{eq_uBT}) - (\ref{eq_vBC}):
\begin{equation}
    \pder[]{y} \Big(\hat{S} - \hat{\Phi}_0 \Big) = \frac{iky}{\sigma} \Big(\hat{S} - \hat{\Phi}_0 \Big) \label{eq_dy}
\end{equation}
Next, we combine Equations (\ref{eq_uBT}), (\ref{eq_uBC}), (\ref{eq_s}), and (\ref{eq_sm}) to eliminate $u_0$, $u_1$, and $s_m$ and obtain the relationship between the meridional function of the saturation moist entropy and the barotropic geopotential:
\begin{align}
    \hat{S} = \lambda \hat{\Phi}_0 \label{trop_disp}
\end{align}
where
\begin{align}
    \lambda &= \frac{i k a_2 + k^2 a_3}{\sigma a_1 + i k a_2 + a_3 k^2} \\
    a_1 &\equiv D(1+C) + (\chi + \sigma)(\gamma \sigma - C) \\
    a_2 &\equiv \alpha(\gamma \sigma + 1) \\
    a_3 &\equiv \gamma \sigma + (G-1)C + G
\end{align}
Combining Equations (\ref{eq_dy}) and (\ref{trop_disp}) gives us the meridional structure $Y$ of the troposphere portion of the $v = 0$ modes:
\begin{equation}
    Y = \exp \Big( \frac{i k}{2\sigma} y^2 \Big) \label{eq_Y_struct}
\end{equation}
which is equivalent in form to the meridional structure of the modes in the rigid-lid case. Furthermore, only solutions with an eastward phase speed satisfy the boundary conditions in $y$.

Next, we move on to solving the portion of the mode that exists in the stratosphere. With the solution form shown in Equation (\ref{eq_strat_sol}), Equations (\ref{eq_strat_U_nd}) - (\ref{eq_strat_T_nd}) reduce to:
\begin{align}
    \sigma \hat{U}_s  + ik \hat{\Phi}_s = 0& \label{strat_lin1} \\
    y \hat{U}_s + \pder[\hat{\Phi}_s]{y} = 0& \label{strat_lin2} \\
    ik \hat{U}_s + \frac{m^2 \sigma}{S} \hat{\Phi}_s = 0  \label{strat_lin3}
\end{align}
where $\hat{W}_s(y) = -\frac{im\sigma}{S} \hat{\Phi}_s(y)$, assuming $ m^2 >> (H / 2 H_s)^2$, which is a vertical short-wave approximation. Equations (\ref{strat_lin1}) and (\ref{strat_lin3}) combine into the well-known dispersion relation for the Kelvin-wave:
\begin{equation}
    \sigma = \pm i \frac{\sqrt{S} k}{m} \label{eq_strat_kelv_dispersion}
\end{equation}
As shown in \citet{yano1991improved}, the associated group velocity is:
\begin{equation}
    c_{g,z} = \frac{k \sqrt{S}}{|m|^4} \Big( \text{Real}(m)^2 - \text{Imag}(m)^2 \Big) \label{kelvin_cgz}
\end{equation}
which indicates that the vertical group velocity increases with zonal wavenumber.

Next, Equations (\ref{strat_lin1}) and (\ref{strat_lin2}) combine to give a meridional structure $Y$ that is equivalent to that of the troposphere shown in Equation (\ref{eq_Y_struct}). For solutions that obey the meridional boundary conditions, or that the mode amplitudes go to zero as $y \rightarrow \pm \infty$, we must have that $\sigma_i < 0$ (equivalent to an eastward phase speed), so we choose the positive root.

We now apply the matching conditions to derive the dispersion relation. The continuity of pressure condition demands that:
\begin{equation}
    \phi_s(x, y, z^* = 1, t) =  \phi_0 - V_1(p_t) s^* \label{eq_phi_match2}
\end{equation}
Equations (\ref{eq_uBT}) and (\ref{eq_omega_cont}) combine to give the vertical velocity at the tropopause in the troposphere, which must be equal to the vertical velocity at the tropopause in the stratosphere:
\begin{equation}
     \omega(p_t) = \frac{k^2}{\sigma} \phi_0 = -\frac{1}{B} w^*_s = \frac{1}{B} \frac{i m \sigma}{S} \hat{\phi}_s \label{eq_w_match2}
\end{equation}
where we note the equivalence of a rigid-lid ($\omega(p_t) = 0$) to the absence of the barotropic mode. Combining with Equations (\ref{eq_phi_tp}), (\ref{trop_disp}), (\ref{eq_strat_kelv_dispersion}), and (\ref{eq_phi_match2}), Equation (\ref{eq_w_match2}) reduces to the dispersion relation:
\begin{equation}
     \underbrace{\sigma a_1 + a_4}_{\text{rigid lid}} + \underbrace{\sigma (k B \sqrt{S})^{-1} (\sigma a_1 + \nu a_4)}_{\text{correction}} = 0 \label{eq_leaky_disp}
\end{equation}
where $\nu = 1 - V_1(p_t)$ and $a_4 = ik a_2 + a_3 k^2$. As shown in the underbraces, Equation (\ref{eq_leaky_disp}) is written in the form of the rigid-lid dispersion relation, plus a quartic-order correction term whose magnitude is inversely propotional to the square root of the stratosphere stratification, $S$. It is clear that in the limit of $S \rightarrow \infty$, the dispersion relation reduces to that of the rigid-lid case. The quartic dispersion relation is solved numerically and the solutions are checked rigorously to satisfy the governing equations, boundary conditions, and matching conditions.

Before examining the solutions of the leaky $v = 0$ modes, we first briefly derive the functional form of $\omega$ in the troposphere, in the case where $v = 0$. We start by taking the time derivative of the non-dimensional continuity equation and substitute the zonal velocity using Equations (\ref{eq_uBT}) and (\ref{eq_uBC}):
\begin{equation}
    \sigma \pder[\omega]{p} = k^2 \phi
\end{equation}
Taking a derivative in pressure allows us to substitute in Equation (\ref{eq_dphi_dp}) and yields:
\begin{equation}
    \sigma \dpder[\omega]{p} = -k^2 \bigg( \pder[T]{p} \bigg)_{s^*} s^{*}
\end{equation}
Integrating in pressure once returns:
\begin{equation}
    \sigma \pder[\omega]{p} = -k^2 s^{*} \overline{T}(p)\Big|_{s^*} + C \label{eq_omega_p_c}
\end{equation}
where $C$ is an integration constant that must be determined through the boundary conditions. We integrate from the surface to the tropopause:
\begin{equation}
    \sigma (\omega(p_t) - \omega(p_s)) = (p_s - p_t)(-k^2 s^* [\overline{T}] +  C)
\end{equation}
which allows us to determine $C$:
\begin{equation}
    C = \sigma \frac{\omega(p_t) - \omega(p_s)}{p_s - p_t} + k^2 s^* [\overline{T}] 
    \label{eq_C}
\end{equation}
Substituting for $C$ into Equation (\ref{eq_omega_p_c}), setting $\omega(p_s) = 0$ as earlier, and integrating once in pressure yields the non-dimensional vertical structure of $\omega$:
\begin{equation}
    \omega(p) = \underbrace{\frac{p_s - p}{p_s - p_t} \omega(p_t)}_{\text{barotropic}} - \underbrace{\frac{k^2}{\sigma} s^* \int_{p_s}^{p_t} V_1 \: dp}_{\text{baroclinic}} 
    \label{eq_omega_decomp}
\end{equation}
As shown in the underbraces, the vertical structure of the pressure vertical velocity is a sum of the barotropic and baroclinic modes. The barotropic component changes linearly with pressure, and the baroclinic mode is zero at the surface/tropopause and peaks in the mid-troposphere. The superposition of the barotropic and baroclinic mode can lead to a vertical tilt in the vertical velocity profile that depends on the phase lag between the two modes. It is worth noting that by definition, the baroclinic mode cannot interact with the stratosphere; in this model, it is only through the excitement of the barotropic mode that waves can radiate energy into the stratosphere.

\begin{figure}[t]
  \includegraphics[width=19pc]{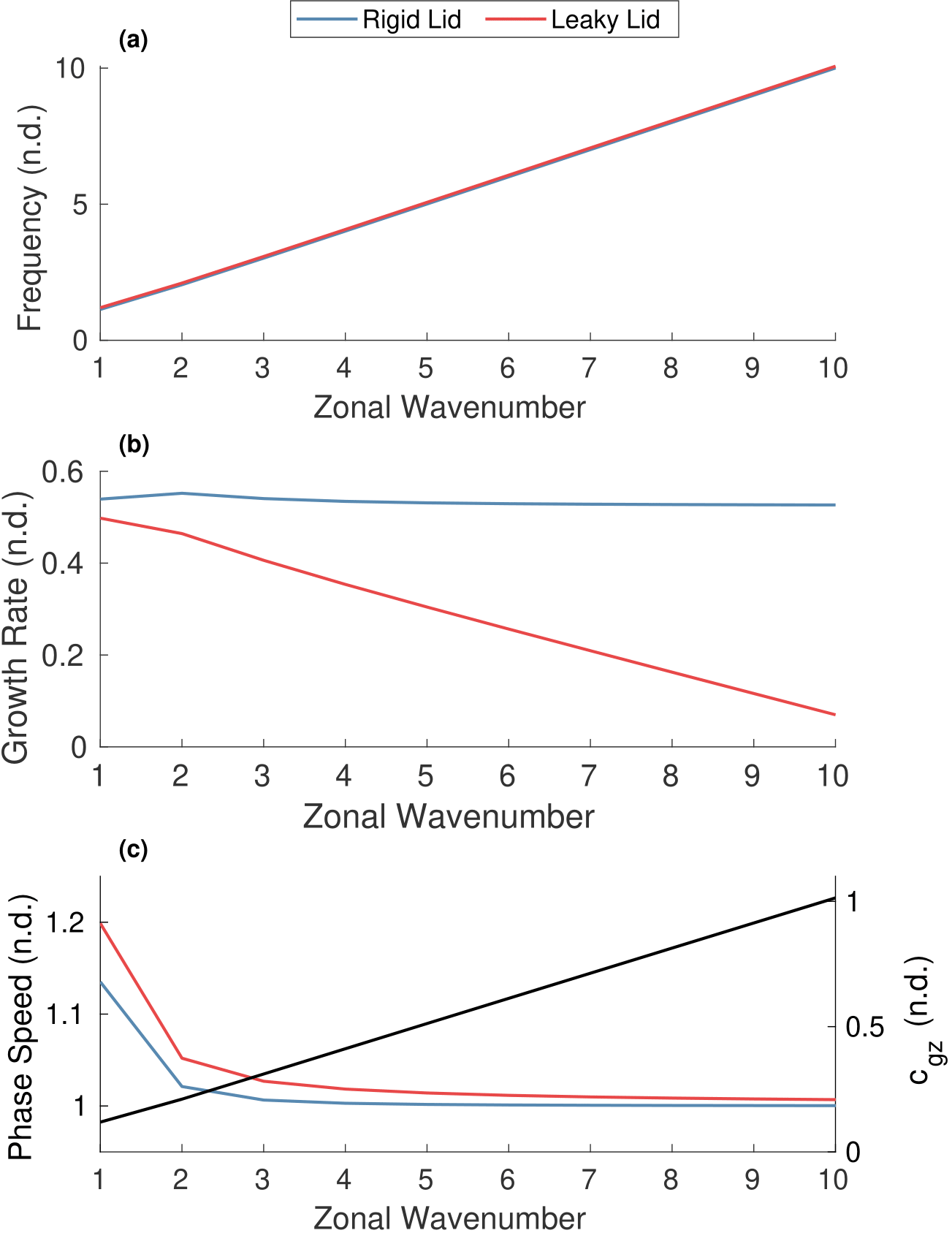}
  \caption{Non-dimensional (a) frequency, (b) growth rate, (c) phase speed and for the fastest growing $v = 0$ mode. Vertical group velocity of the leaky mode is shown in black in the bottom panel, while rigid-lid solutions are shown in blue and leaky-lid solutions shown in red. Non-dimensional parameters are $\alpha = 1.5$, $\chi = 0.5$, $C = 0$, $\gamma = 2$, $D = 0.5$, $G = 0.1$, $S = 100$.} \label{f1}
\end{figure}

Figure \ref{f1} compares the non-dimensional frequency, growth rates, phase speeds, and vertical group velocity of the leaky $v = 0$ modes to those of the rigid-lid modes, with non-dimensional coefficients $\alpha = 1.5$, $\chi = 0.5$, $C = 0$, $\gamma = 2$, $D = 0.5$, $G = 0.1$, $S = 100$. These parameters somewhat reflect Earth-like conditions in the tropics, with the exception of no cloud radiation interaction, and were specifically chosen to examine the branch of solutions that closely resemble the classical Kelvin-wave solutions but are instead amplified through the WISHE feedback. Figure \ref{f1}a shows that the frequencies of the leaky waves are only slightly larger than their rigid-lid counterparts, and as such the phase speeds are slightly faster (Figure \ref{f1}c). The modification of the growth rates tell a much different story: Figure \ref{f1}b shows strong damping of the growth rate of smaller-scale waves, shifting the power spectrum towards larger scale waves, which is consistent with the results of \citet{yano1991improved}. This reddening of the power spectrum can be physically explained through the large vertical group velocities of the smaller scale waves, as shown in Figure \ref{f1}c and indicated in Equation (\ref{kelvin_cgz}). The smaller scale waves propagate their energy very quickly into the stratosphere, dampening their growth rate. For instance, the $k=10$ wave has a vertical group velocity that nearly exceeds its zonal phase speed.

\begin{figure*}
  \includegraphics[width=39pc]{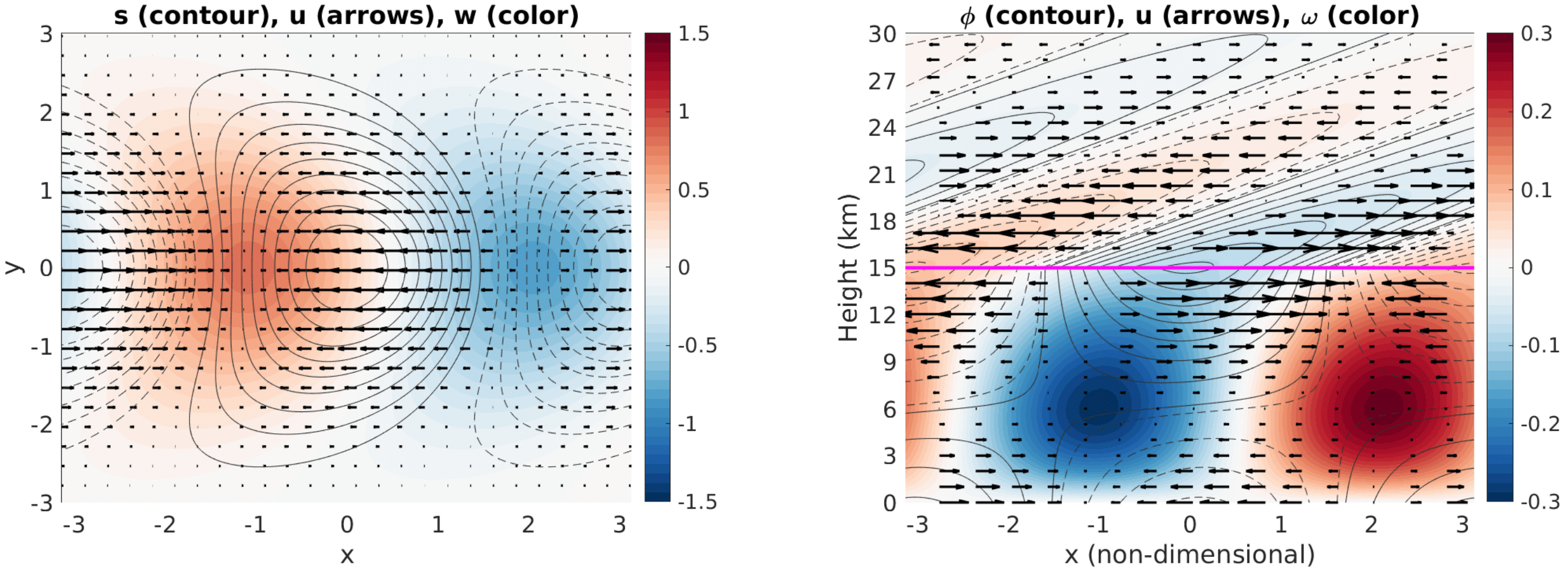}
  \caption{Eigenfunction of the $v = 0$, $k = 1$ mode corresponding to the parameters detailed in Figure \ref{f1}, on a (left) horizontal cross-section at the surface, and (right) vertical cross-section centered on the equator. Contours indicate the saturation entropy or geopotential, where solid (dashed) lines indicate positive (negative) perturbations. Arrows indicate zonal velocity perturbations, and color shading indicates vertical velocity or $\omega$ perturbations, where indicated. Magenta line outlines the tropopause.} \label{f2}
\end{figure*}

Figure \ref{f2} shows a summary of the $k = 1$ eigenfunction using identical non-dimensional parameters as detailed in Figure \ref{f1}. The horizontal cross-section at the surface shows a Kelvin-wave pattern with surface easterlies maximizing east of the maximum temperature anomaly. Vertical velocity maximizes west of the peak temperature anomalies, and the WISHE feedback is responsible for wave growth and enhanced eastward propagation. The horizontal structure is not significantly modified from that of the rigid-lid. The vertical structure of the mode exhibits a bit more complexity. Unlike the vertical structure of the rigid-lid model, which is purely baroclinic, the superposition of the barotropic and baroclinic modes leads to an eastward vertical tilt, as shown in Figure \ref{f2} and Equation (\ref{eq_omega_decomp}). The mode has a downward phase propagating, eastward tilted component in the stratosphere, which is consistent with a Kelvin-wave that has upward vertical energy propagation. As indicated by the dispersion relation, the vertical wavelength in the stratosphere is controlled by the horizontal wavenumber and stratification; a stronger stratification or shorter horizontal wavelength decreases the vertical wavelength.

\begin{figure}{t}
 \includegraphics[width=19pc]{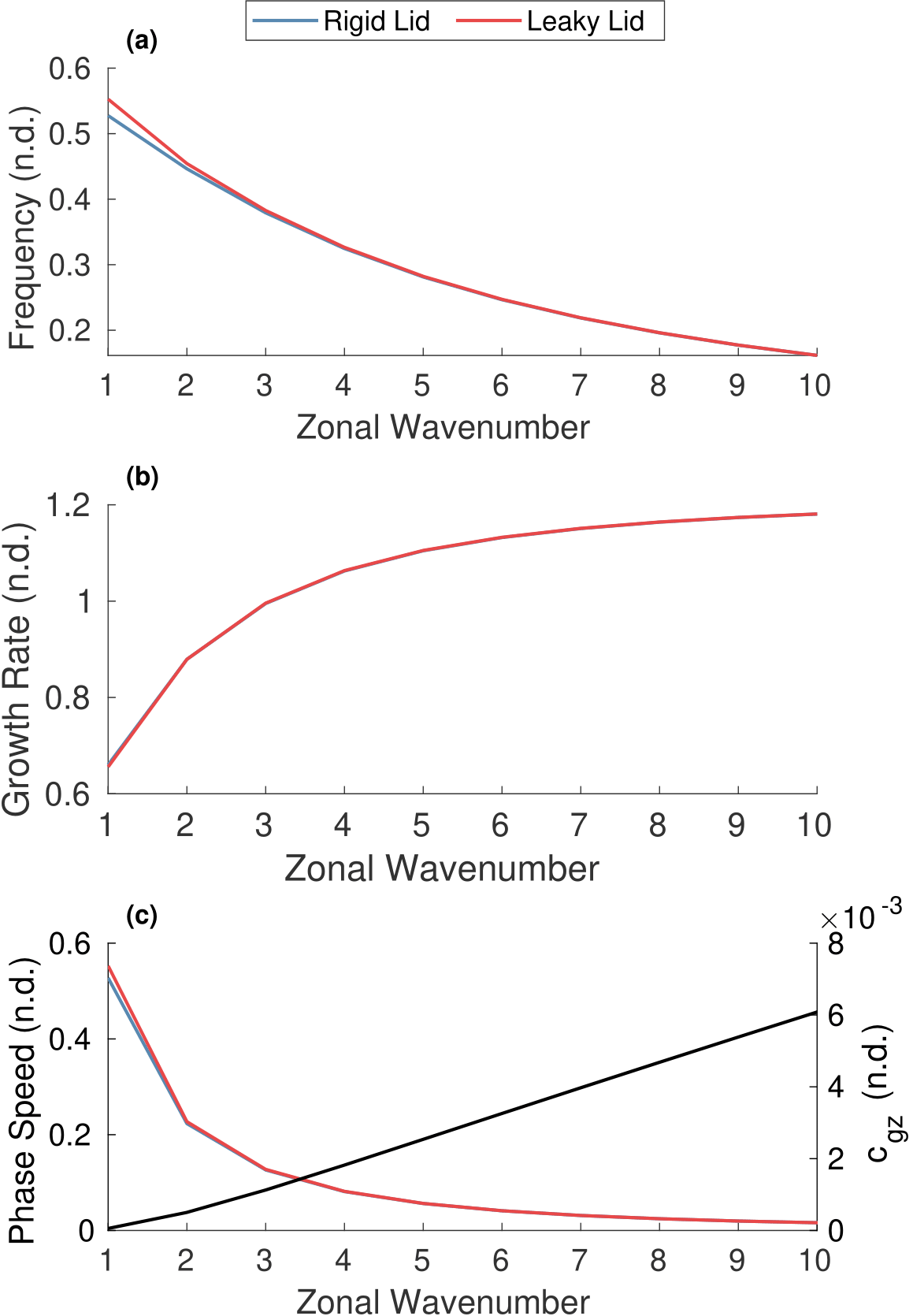}
  \caption{Analogous to Figure \ref{f1}, but for the fastest-growing mode with non-dimensional parameters $\alpha = 1$, $\chi = 1$, $C = 2.5$, $\gamma = 2$, $D = 1$, $G = 0.02$, $=100$.} \label{f3}
\end{figure}

The tropospheric system also allows for solutions of significantly slower propagating modes when cloud-radiation interaction is turned on, as shown in KE18 and E20. We select the non-dimensional parameters $\alpha = 1$, $\chi = 1$, $C = 2.5$, $\gamma = 2$, $D = 1$, $G = 0.02$, $S = 100$, in order to obtain the slow modes as the fastest growing solutions. Note that since we eliminated horizontal diffusion of MSE as compared to the model described in KE18, the growth rates no longer peak at low wavenumbers.  Figure \ref{f3} shows the frequency, growth rate, phase speed, and vertical group velocities for the $v = 0$ slow modes that can interact with the stratosphere. Aside from a small modification of the wave properties at the highest wave numbers, these slow modes are not significantly affected by the presence of a stratosphere. Their vertical group velocities are almost negligibly small, and the corresponding vertical wavelengths in the stratosphere are extremely short. These slow modes are trapped in the troposphere and do not leak much energy into the stratosphere. Thus, these results suggest that for slow propagating modes, such as the MJO, a rigid-lid assumption is an accurate approximation with regards to the growth rates.

Next, we investigate the barotropic mode magnitude and phase tilt across a range of non-dimensional parameters. The phase lag and amplitude relationship between the baroclinic and barotropic zonal velocities is:
\begin{equation}
    u_0 = \Big( \frac{k B \sqrt{S}}{\sigma \lambda} - \nu \Big) u_1 \label{eq_phase_lag}
\end{equation}
Equation (\ref{eq_phase_lag}) shows that the amplitude of the barotropic mode decreases with increasing zonal wavenumber and stratospheric stratification. For Earth-like parameters, the amplitude of the barotropic mode for the $k=1$ solution is around an order of magnitude smaller than the amplitude of the baroclinic mode. The phase lag is determined by the complex coefficient in parentheses in Equation (\ref{eq_phase_lag}).

\begin{figure*}
 \includegraphics[width=39pc]{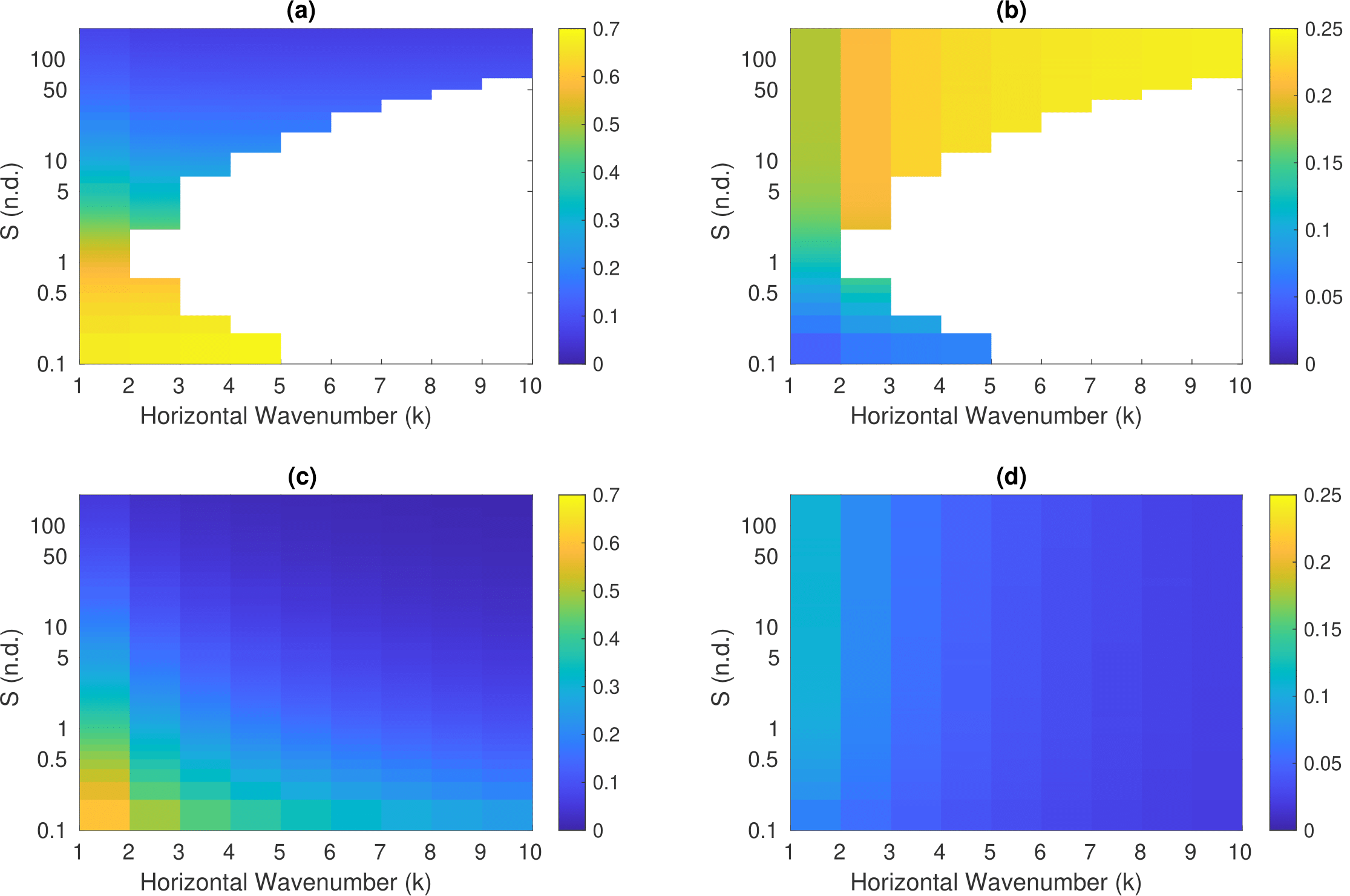}
 \caption{(a, c) Barotropic zonal velocity to total zonal velocity ratio [$u_0$ / ($u_0 + u_1$)], and (b, d) phase lead of the barotropic mode with respect to the baroclinic mode, in radians, as functions of stratosphere stratification, $S$, and horizontal wavenumber. (a, b) Top row  generated using the non-dimensional parameters described in Figure \ref{f1} (WISHE-modified Kelvin waves), and (c, d) bottom row generated using the non-dimensional parameters described in Figure \ref{f3} (slow $v = 0$ modes). White area indicates non-growing modes.} \label{f4}
\end{figure*}

Figure \ref{f4}a and Figure \ref{f4}c show the barotropic zonal velocity to total zonal velocity ratio, which we define as $u_0 / (u_0 + u_1)$, as well as the phase lead of the barotropic mode to the baroclinic mode, for a wide range of stratosphere stratification $S$ (an Earth-like range of stratosphere stratification is $S \approx 20 - 200$). In general, we observe decreasing amplitude of the barotropic mode with increasing stratosphere stratification and zonal wavenumber, as suggested by Equation (\ref{eq_phase_lag}). For both the classical WISHE-driven Kelvin waves (Figure \ref{f4}, top row) and radiatively destabilized slow modes (Figure \ref{f4}, bottom row), the barotropic mode is around an order of magnitude smaller than the baroclinic mode for Earth-like parameters. Furthermore, the barotropic mode leads the baroclinic mode by approximately 0.1-0.2 radians. The phase lead of the barotropic mode is always positive, indicating that the eastward phase tilt shown in Figure \ref{f2} is robust across a wide-range of scales and $S$. In addition, an exploration of the non-dimensional parameter space did not find any solutions where the barotropic mode lagged the baroclinic mode (not shown). This result is consistent with physical intuition. For the WISHE-driven $v = 0$ modes, the baroclinic easterlies peak east of the maximum temperature perturbation. If the barotropic zonal velocities are instead westward of the baroclinic zonal velocities, then the total surface easterlies could peak west of the maximum temperature anomalies, which is inconsistent with an WISHE-driven eastward propagating mode. Note, this is inconsistent with observational analyses of convectively coupled Kelvin waves, in which a westward tilt with height is observed~\citep{straub2002observations}. This may indicate the importance of the second baroclinic mode for Kelvin waves~\citep{mapes2000convective, kuang2008modeling}.

Finally, it is worth noting that the $u_0$ magnitude is larger than the $u_1$ magnitude for extremely leaky stratospheres (Figure \ref{f4}); though this may not be relevant to Earth's atmosphere, a barotropic mode that exceeds its baroclinic counterpart leads to top-heavy vertical velocity profiles and fast propagation into the stratosphere.

\subsubsection{Higher order modes}

Without the $v = 0$ approximation, the solution set is extremely complex. In the rigid-lid case, the meridional structure of the eigenmodes of the full equations are parabolic cylinder functions of degree $n$, where $n$ is the order of the Hermite polynomial. A brief mathematical analysis of the (intractable) leaky-lid solutions suggests that the meridional structure functions of the linear solutions are sums of parabolic cylinder functions. Note that by convention, we denote $n = -1$ solutions as those where $v = 0$.

For these higher order modes, we instead solve the linear problem using numerical code, which is only appropriate for finding growing modes. The troposphere system [Equations (\ref{eq_uBT}) - (\ref{eq_vBC}), (\ref{eq_s}) - (\ref{eq_sm})] is discretized in $y$, while the stratosphere system [Equations (\ref{eq_strat_U_nd}) - (\ref{eq_strat_T_nd})] are discretized in $y$ and $z$. Solutions are assumed to have a zonal structure of the form $\exp(ikx)$. In order to integrate the linear system in time, the stratosphere equations need to be transformed to a set of linear prognostic equations. The mass continuity equation [Equation (\ref{eq_strat_cont_nd})] is first integrated from the lower boundary in $z^*$, to obtain:
\begin{strip}
\begin{equation}
    \rho_s w^*_s(y, z^*) = w^*_s(y, z^* = 1) - \int_{z^* = 1}^{z} \Big[ \rho_s \Big( i k u_s(y, z^*) + \\ \pder[]{y} v_s(y, z^*) \Big) \Big] dz^*
\end{equation}
\end{strip}
where the tropopause velocity $w^*(z^* = 1)$ is equal to the vertical velocity at the tropopause in the troposphere equations, as required from the vertical velocity matching condition. Given the vertical velocity, Equation (\ref{eq_strat_T_nd}) can be used to calculate the geopotential:
\begin{equation}
    \pder[]{t} \pder[]{z} \phi(y, z^*) = - w^*(y, z) S
\end{equation}
Integrating from the upper boundary downwards gives the prognostic equation for $\phi_s$:
\begin{equation}
    \pder[]{t} \phi_s(y, z^*) = - \int^{z}_\infty w^*_s(y, z^*) S \: dz^* + C_0
\end{equation}
where $C_0$ must be determined by the upper boundary condition. An upwards wave-radiation boundary condition could be applied to determine $C_0$, as in \citet{moskowitz2000analysis}, but such a condition does not exist for the slower modes that have interactive cloud radiation and no clear analog in the stratosphere. While it is not necessary that $\phi_s = 0$ as $z \rightarrow \infty$ (as long as the energy density goes to zero), we have included a Newtonian damping in a sizeable layer extending from the upper boundary, and as such, we set $C_0 = 0$. Another advantage of the Newtonian damping in the upper part of the domain is to eliminate any nonphysical downward propagating modes. For realistic values of $S$, however, the amplitude of the modes typically decay to zero at the top of the numerical domain. Finally, $\phi_s(y, z^* = 1)$ is connected to the troposphere through the continuity of pressure matching condition [Equation (\ref{eq_phi_match2})]. 

Spatial derivatives are approximated using fourth-order central differences, and the system is forward time-stepped using fourth order Runge-Kutta. After specifying non-dimensional parameters, the corresponding rigid-lid solution is used to initialize the troposphere domain, while the stratosphere is initialized at rest. Since the unbalanced wave must undergo rapid gravity-wave adjustment, several dampening mechanisms are used to eliminate spurious gravity-wave energy. A spectral filter is applied at each time step to eliminate small-scale noise. A strong sponge-layer is included along the edges and top of the domain to eliminate reflection of gravity waves, downward propagating waves, and spurious noise. For details on the mathematical form of the full numerical system, see Appendix B. The system is integrated for a long period of time, during which the domain is periodically rescaled by a constant to prevent numerical overflow. We then isolate the growing mode of interest and infer both the complex growth rate and meridional/vertical structures. Although we initially assume that the damping at the upper boundary does not significantly affect the stratospheric solution, the corresponding growth rate and the meridional structures of all prognostic variables are rigorously checked to satisfy the governing equations, boundary conditions, and matching conditions. As a partial test of correctness of the numerical code, the numerical solutions for the $v = 0$ ($n = -1$) modes were cross-referenced with the $v = 0$ analytic solutions.

The numerical solution is robust for linear solutions that are the fastest growing equatorially symmetric or asymmetric mode. However, for solutions that are not the fastest growing mode, initial unbalanced energy and/or numerical error that projects onto the fastest growing mode symmetric/asymmetric mode will cause the slower growing mode to be overtaken by the fastest growing mode before the domain energy is concentrated solely in the mode of interest. For instance, the second fastest growing equatorially symmetric mode can be overtaken by the fastest growing equatorially symmetric mode. In these cases, the meridional/vertical structure of the fastest growing mode is isolated, projected onto the domain output, and removed from the fields. If the second fastest growing mode does not have a weak growth rate, the remaining, filtered fields will contain the mode of interest for a long enough period of time to infer the complex growth rate and structure. 

\begin{figure*}
  \includegraphics[width=39pc]{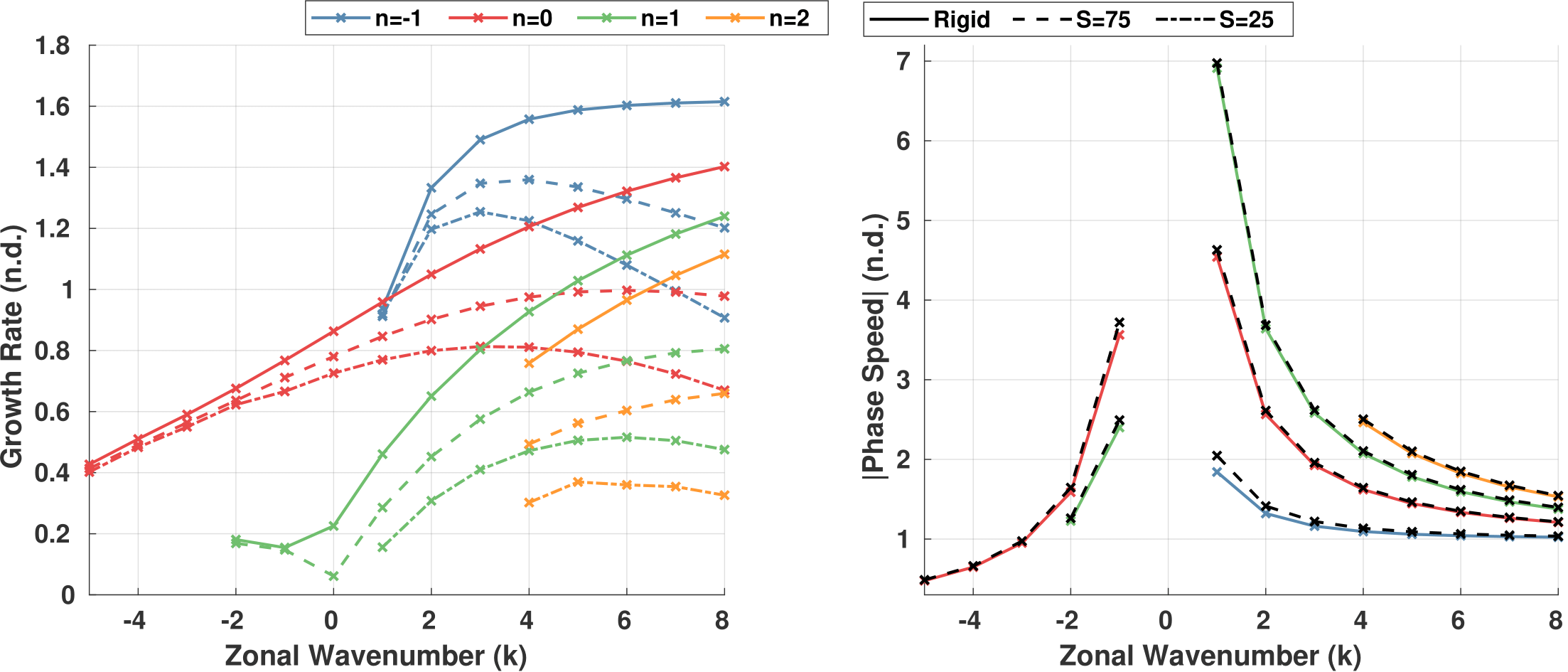}
  \caption{(Left) Non-dimensional growth rate for the $n = -1, 0, 1, 2$ modes of the (solid) rigid-lid system, the (dashed) leaky-lid system with $S = 75$, and the (dot-dashed) leaky-lid system with $S = 25$. (Right) Non-dimensional phase speed for the same modes, in the (solid) rigid-lid and (black-dashed) leaky-lid system with $S = 75$. $k <= 3$ for $n = 2$, and $k <= 0$ for $n = 1$ for $S = 25$ solutions are omitted since they do not grow rapidly enough to infer the complex growth rate. Non-dimensional parameters selected for these modes are $\alpha = 3.5$, $\chi = 0.5$, $C = 0$, $\gamma = 1$, $D = 2.5$, $G = 0.25$, $\delta_x = 15$.} \label{f5}
\end{figure*}

\paragraph{WISHE classical modes}
We first compare the leaky wave solutions to the rigid-lid solutions for the WISHE-destabilized classical modes of the equatorial waveguide. This is done by choosing $\alpha = 3.5$, $\chi = 0.5$, $C = 0$, $\gamma = 1$, $D = 2.5$, $G = 0.25$, $\delta_x = 15$, $S = 75$; note that $C = 0$ eliminates the slower propagating modes observed in KE18 and E20. Figure \ref{f5} shows the non-dimensional growth rates and phase speeds for select meridional orders $n = -1, 0, 1, 2$ leaky equatorial waves. The growth rates of the Kelvin, westward and eastward mixed Rossby-gravity, and inertia-gravity waves are clearly dampened in relation to their rigid-lid counterparts. The strength of dampening, in a percent relative sense, grows stronger with increasing wavenumber for the eastward modes. This is in contrast to the almost negligible dampening of the westward propagating modes as $|k|$ increases. The growth rates of the westward propagating $n =0$ and $n =1$ waves are not strongly affected by the stratosphere. The phase speeds of the leaky-waves are almost negligibly faster than their rigid-lid counterparts, with the largest, though still slight, modifications observed for the $n = -1$ solutions. The growth rates for smaller scale eastward propagating waves are further dampened with reduced stratification in the stratosphere (compare $S = 75$ to $S = 25$). The effect is quite pronounced; in the case shown in Figure \ref{f5}, the smallest scale rigid-lid modes possess the largest growth rates, but imposing a leaky stratosphere greatly shifts the growth rates towards larger scales. For the $S = 25$ case, the growth rate peaks at $k = 3$ for the $n = -1, 0$ modes, and $k \approx 5$ for the $n = 1, 2$ modes. These results are consistent with the qualitative behavior shown in the mathematical analysis of the leaky $v = 0$ modes. As such, the solutions suggest that the stratosphere acts as a reddener of the equatorial power spectrum, especially for eastward propagating waves.

\paragraph{Slow modes}
We next switch to a non-dimensional parameter set in which cloud radiative feedbacks are turned on, such that the slow modes analyzed in E20 are the fastest growing modes. In particular, we choose $\alpha = 1$, $\chi = 1$, $C = 2.5$, $\gamma = 2$, $D = 1$, $G = 0.02$, $\delta_x = 30$, $S=100$. Although the westward propagating slow modes do grow in time under the rigid-lid model, they take too long to reach a steady state and are quickly overtaken by the eastward propagating modes. We thus focus on the eastward propagating slow modes, which have a much clearer analog in the real atmosphere (i.e. the MJO).

\begin{figure*}
  \includegraphics[width=39pc]{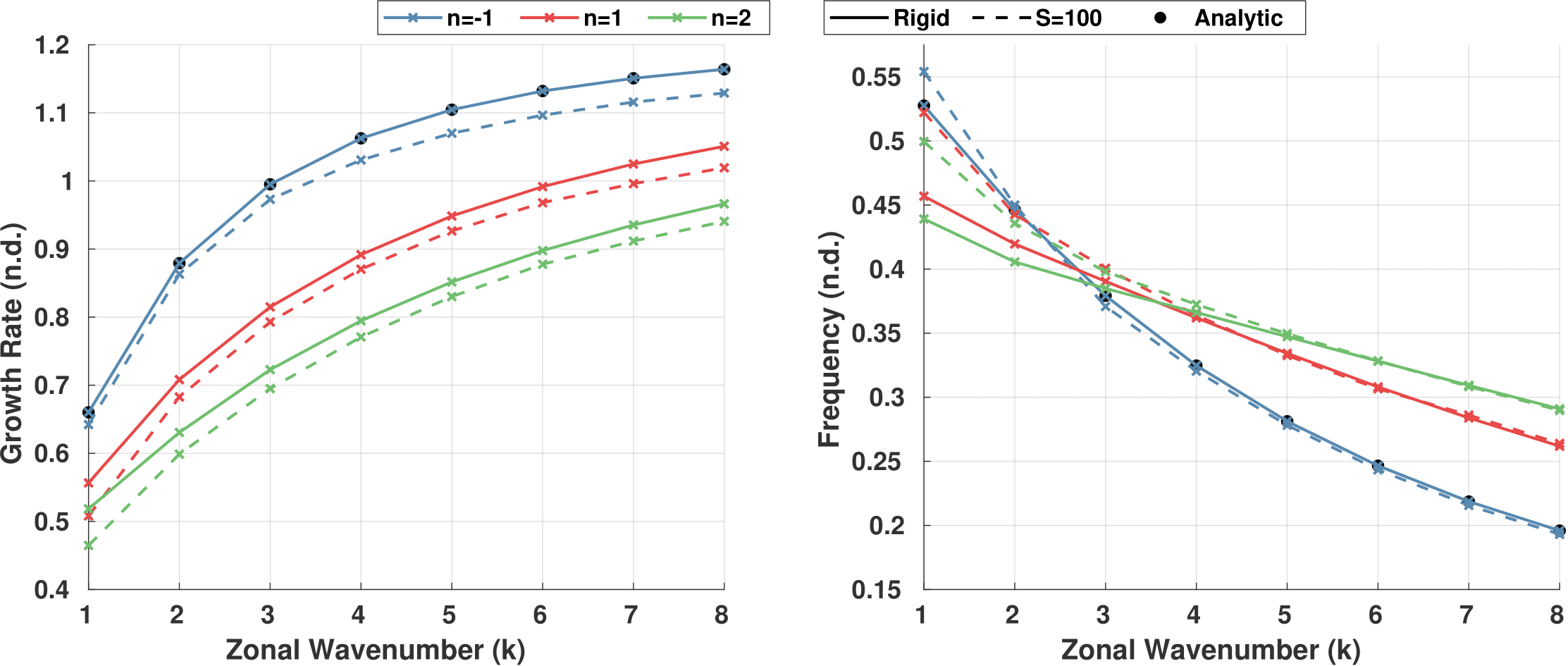}
  \caption{(Left) Non-dimensional growth rate for the $n = -1, 1, 2$, eastward propagating slow modes of the (solid) rigid-lid system, and the (dashed) leaky-lid system with $S = 100$ inferred from the numerical solution. (Right) Non-dimensional frequency for the same modes, in the (solid) rigid-lid and (dashed) leaky-lid system with $S = 100$. Black dots indicate the analytic solutions of the growth rate and frequencies of the $n = -1$ slow modes. All $n=0$ solutions are not growing. Non-dimensional parameters selected for these modes are $\alpha = 1$, $\chi = 1$, $C = 2.5$, $\gamma = 2$, $D = 1$, $G = 0.02$, $\delta_x = 30$.} \label{f6}
\end{figure*}

Figure \ref{f6} shows the non-dimensional growth rates and frequencies for the eastward propagating $n = -1, 1, 2$ slow modes. We first compare the numerical $n = -1$ solutions to the analytic solutions (Figure \ref{f6}, black dots). We observe that the numerical model slightly dampens the growth rate, and only negligibly modifies the frequencies. This slight damping could be attributed to the spectral filter and/or numerical error. Assuming that the numerical error affects the wave characteristics of the $n = 1, 2$ slow modes the same way, we can infer that like the growth rates and frequencies of the leaky $n = -1$ slow modes, the growth rates and phase speeds of the higher order slow modes are not affected by the presence of the stratosphere.

\begin{figure*}
  \includegraphics[width=39pc]{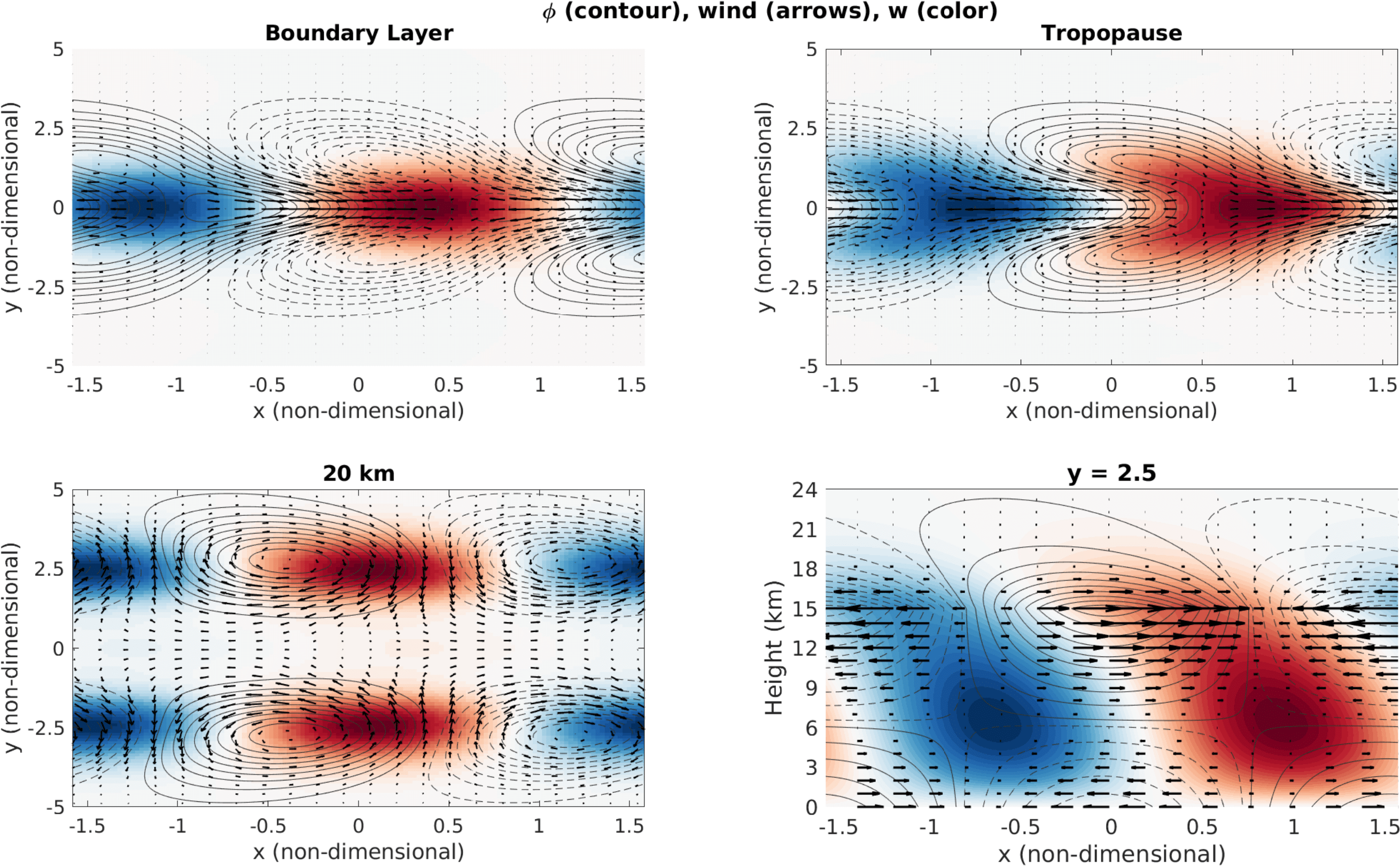}
  \caption{Eigenfunction of the $n = 1$, $k = 2$ mode corresponding to the parameters detailed in Figure \ref{f6}, on a horizontal cross-section at the (top left) the surface, (top right) tropopause [15 km], (bottom left) 18 km. Vertical cross section eigenfuction at $y = 2.5$ is shown on the bottom right. Contours indicate the geopotential, where solid (dashed) lines indicate positive (negative) perturbations. Arrows indicate wind perturbations, and color shading indicate vertical velocity perturbations at the level indicated, except for the boundary layer cross-section, where color shading indicates mid-level vertical velocity.} \label{f7}
\end{figure*}

Perhaps of greater interest is how the slow modes interact with the stratosphere through the barotropic mode. We focus on the eastward propagating $n = 1$, $k = 2$ mode which has a horizontal structure that closely resembles that of the observed MJO~\citep{emanuel2020slow}. Figure \ref{f7} shows the horizontal and vertical structure of the $n = 1$, $k = 2$ mode, where we observe surface cyclonic gyres westward and poleward of the area of maximum ascent. Strong westerlies are also observed westward of the maximum ascent region. The boundary layer horizontal structure very closely resembles that of the rigid-lid solution (not shown). At the tropopause (Figure \ref{f7}, top right), the horizontal structure is almost nearly the opposite of the horizontal structure at the surface, indicating the prominence of the first baroclinic mode. However, the vertical velocity at the tropopause is non-zero, and the pattern extends poleward into the cyclonic/anti-cyclonic gyres, an indication of the presence of the barotropic mode. If we move further up into the stratosphere, at around 20 km (Figure \ref{f7}, bottom left), the signature of the equatorial portion of the tropospheric wave disappears, and the poleward cyclonic/anti-cyclonic gyres become the prominent pattern. These gyres have a westward tilt with height, as indicated in the vertical cross section at $y = 2.5$ (Figure \ref{f7}, bottom right), though their amplitudes decay exponentially with height and are much smaller with respect to the amplitudes of the corresponding gyres in the troposphere. At the equator, the $n = 1$, $k = 2$ slow mode has a very weak eastward tilt with height (not shown), since the barotropic mode leads the baroclinic mode. Like the $v = 0$ mode shown in Figure \ref{f2}, in the stratosphere, the slow mode also tilts eastward with height at the equator (not shown). This is consistent with the stratospheric eastward tilt with height of the zonal wind on the equator of observational MJO composites [see fig. 3 of \citet{kiladis2005zonal}].

\subsection{Frictionally modified modes}

In this section, we consider how surface friction can also excite the barotropic mode. First, we simplify the linear system in the rigid-lid limit, but now with non-zero surface friction. From Equation (\ref{eq_omega_cont}), if $\omega(p_t) = 0$, it follows that the barotropic mode is non-divergent. However, the solution cannot be solved in terms of a non-divergent stream function $\psi_0$, since $\delta_x$ scales the meridional momentum equation but not the mass continuity equation. Instead, we take the divergence of Equations (\ref{eq_uBT}) and (\ref{eq_vBT}), yielding:
\begin{strip}
\begin{equation}
      \Big( \delta_x \dpder[]{y} - k^2 \Big) \phi_0 =  y \Big( i k v_0 - \delta_x \pder[u_0]{y} \Big) - \delta_x u_0  - 2 F i k (u_0 + u_1) - F \pder[]{y} \Big( v_0 + u_1 \Big) \label{bt_div}
\end{equation}
\end{strip}
From Equation (\ref{bt_div}), we see that a barotropic mode at rest cannot be excited if there is no surface friction ($F = 0$). Thus, surface friction acts to couple the barotropic mode with the baroclinic mode; since the forcing is primarily in the first baroclinic mode; it is the first baroclinic mode that excites the barotropic mode. 

The barotropic geopotential $\phi_0$ is solved by inverting Equation (\ref{bt_div}). Equations (\ref{eq_uBT}) - (\ref{eq_vBC}), (\ref{eq_s}), (\ref{eq_sm}) form the complete linear system. The system is solved numerically by discretizing in $y$. Spatial derivatives are again approximated using fourth-order central differences, and the system is forward time-stepped using fourth order Runge-Kutta. As before, the frictionless, rigid-lid solution is used to initialize the domain, after which we integrate the system for a long period of time and then isolate the growing mode of interest.

\begin{figure*}
  \includegraphics[width=39pc]{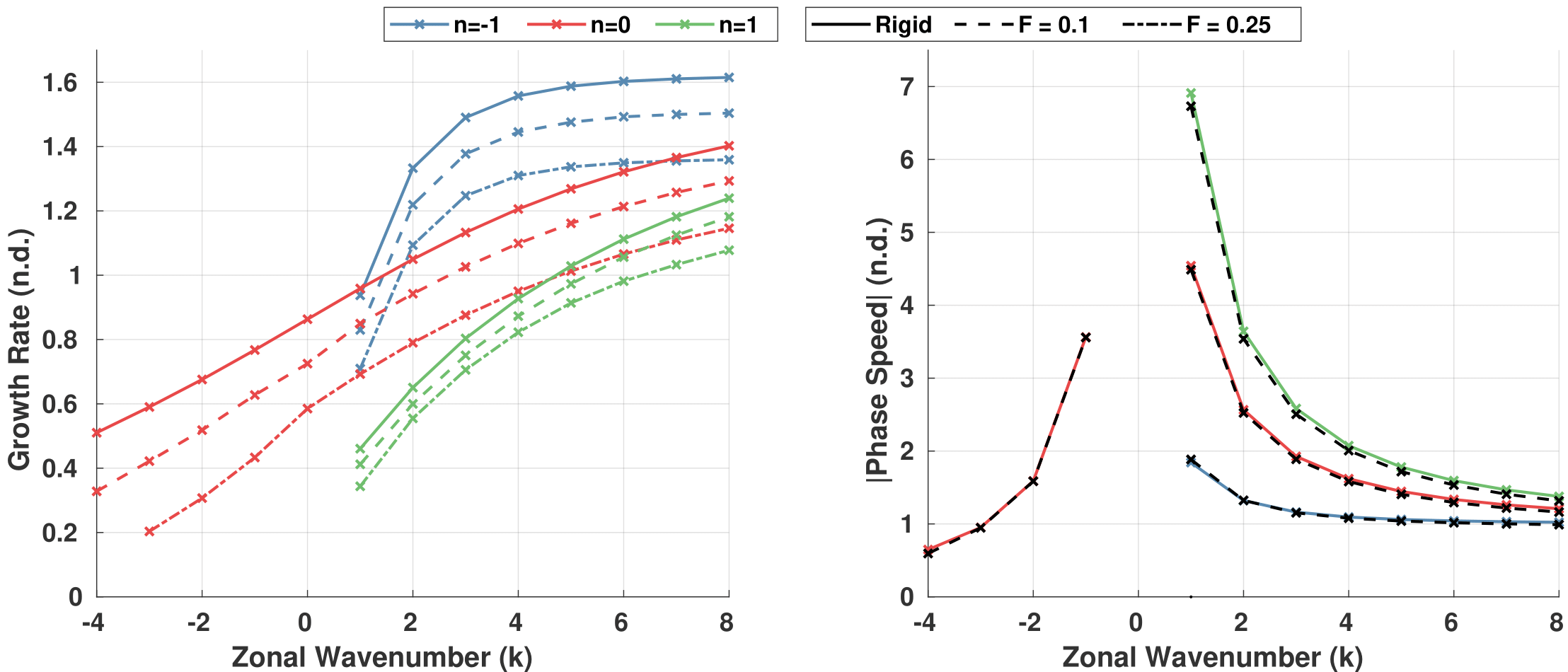}
  \caption{(Left) Non-dimensional growth rate for the $n = -1, 0, 1$ modes of the (solid) rigid-lid system, the (dashed) surface friction + rigid-lid system with $F = 0.1$, and the (dot-dashed) surface friction + rigid-lid system $F = 0.25$. $n = -1$ modes differ from the traditional definition of $v = 0$, in that the meridional wind is allowed to be non-zero but small. (Right) Non-dimensional phase speed for the same modes, in the (solid) rigid-lid and (black-dashed) surface friction + rigid-lid system with $F = 0.1$. All $n = 2$, and $n = 1$ for $k <= 0$ solutions are omitted since they do not grow rapidly enough to infer the complex growth rate. Non-dimensional parameters selected for these modes are $\alpha = 3.5$, $\chi = 0.5$, $C = 0$, $\gamma = 1$, $D = 2.5$, $G = 0.25$, $\delta_x = 15$.} \label{f8}
\end{figure*}

First, we investigate the complex growth rates as a function of the non-dimensional surface friction, $F$, for an Earth-like range is $F \approx 0.1 - 0.4$. Figure \ref{f8} shows the growth rate of the $n = -1$, $n = 0$, and $n = 1$ modes for the same parameter choices as in Figure \ref{f5}, which selects for WISHE-modified Matsuno modes. It is important to note that the $n = -1$ modes in this section differ from the traditional definition of $v = 0$, in that the meridional wind is allowed to be non-zero but small, since $v_0 = v_1 = 0$ solutions do not exist under a fixed surface and rigid-lid assumption. Figure \ref{f8} indicates that the surface friction acts as a damping effect on all wavelengths, and the strength of the damping is nearly constant across all wavelength. This is expected as the aerodynamic drag law acts on a fixed damping time scale. The phase speeds are not significantly changed from the phase speeds of the rigid-lid modes.

\begin{figure*}
  \includegraphics[width=39pc]{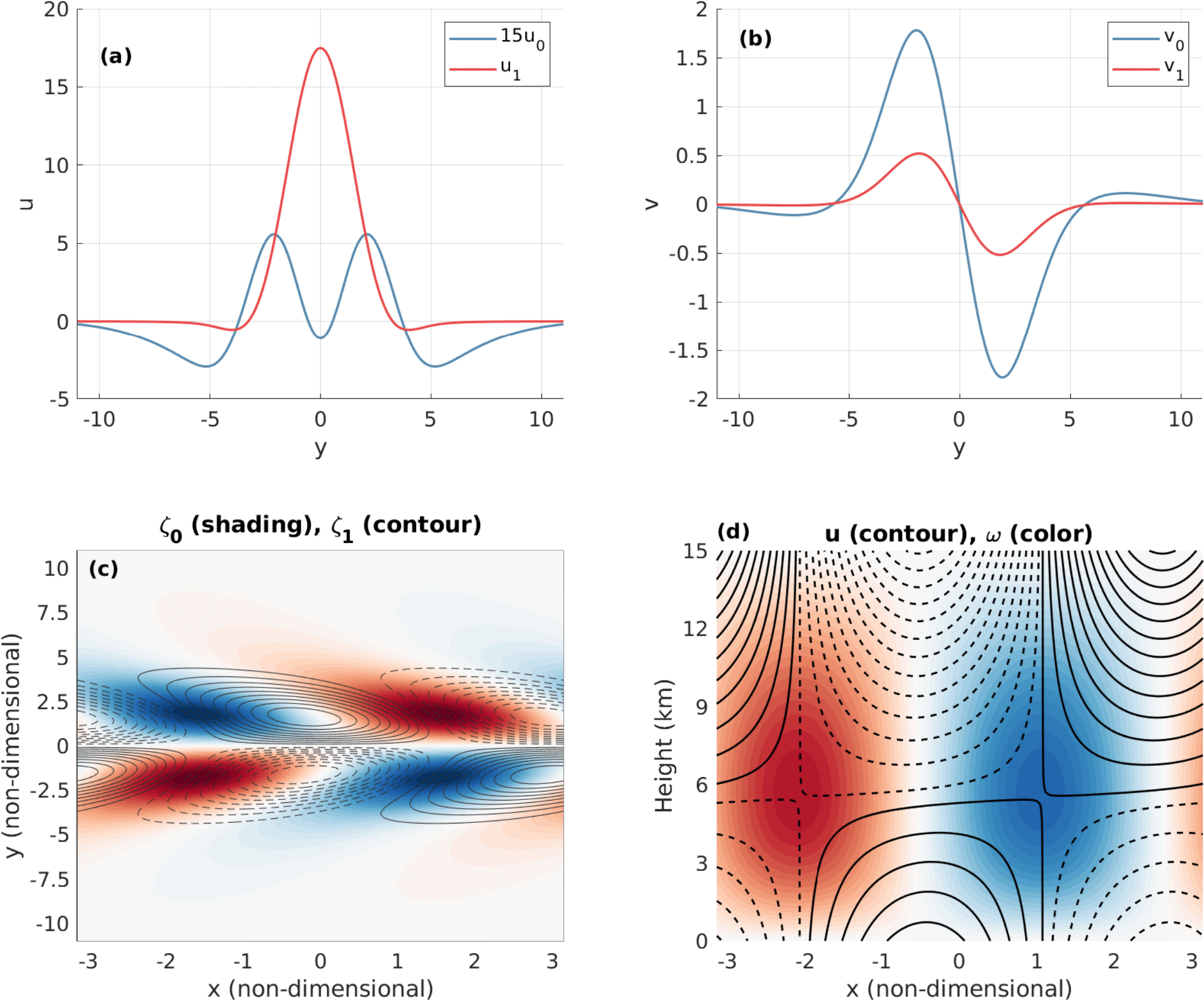}
  \caption{(a) Meridional structure of 15 times the barotropic zonal wind and the baroclinic zonal wind for the $n = -1$, $k = 1$ mode with $F = 0.25$ and the parameter set chosen in Figure \ref{f8}. (b) Same as (a) but for the barotropic and baroclinic meridional wind. (c) Horizontal structure of the boundary layer (shading) barotropic vorticity and (contour) baroclinic vorticity. (d) Equatorial vertical cross sections of the (shading) pressure vertical velocity and (contours) zonal wind perturbations for the same mode. Solid (dashed) contours are positive (negative) zonal wind anomalies.} \label{f9}
\end{figure*}

As illustrated in Equation (\ref{bt_div}), surface friction can lead to an excitation of the barotropic mode. Figure \ref{f9}a and Figure \ref{f9}b show the meridional structure of the $n = -1$, $k = 1$ barotropic zonal and meridional velocities as compared to their baroclinic counterparts. Like the leaky-lid case, the barotropic zonal wind magnitude is around an order of magnitude smaller than the baroclinic mode. However, while the baroclinic mode magnitudes are primarily confined close to the equator (around $y = -5$ to $y = 5$), the barotropic mode wind velocities have long tails that extend far away from the equator. These long tails can be understood from the barotropic mode vorticity equation, which can be solved for by taking the curl of Equations (\ref{eq_uBT}) and (\ref{eq_vBT}):
\begin{equation}
    \sigma \zeta_0 + v_0 = - F \bigg[ \frac{1}{\delta_x} \pder[v_0]{x} - 2 \pder[u_0]{y}  + \frac{1}{\delta_x} \pder[v_1]{x} - 2 \pder[u_1]{y} \bigg]  \label{eq_vortBT}
\end{equation}
where the barotropic mode vorticity $\hat{\zeta}_0 = \hat{\nabla}_H \times \Vec{v}_0$, and $\hat{\nabla}_H = [\frac{1}{\delta_x} \pder[]{x}, \pder[]{y}]$. From Equation (\ref{eq_vortBT}), it is evident that the barotropic mode vorticity can be excited by not only the baroclinic mode vorticity, but also by itself. The generation of barotropic vorticity far from the equator [see Figure \ref{f9}c] leads to non-negligible barotropic winds far from the equator. While further analysis of this effect is out of the scope of this study, the barotropic mode might be an important teleconnection mechanism between the tropics and the extratropics~\citep{horel1981planetary}. An important caveat to note, however, is that a rigid-lid barotropic mode is completely trapped in the troposphere; the addition of a leaky stratosphere may limit the poleward extent of the barotropic mode, as will be examined in the next section.

Under a rigid lid the non-divergent barotropic mode cannot be associated with vertical velocity perturbations, though a vertical tilt can still exist in the horizontal wind fields from the superposition of the barotropic and baroclinic horizontal winds. Figure \ref{f9}d shows the vertical cross section on the equator of the $n = -1$, $k = 1$ mode. The vertical structure of the vertical velocity is only first baroclinic, despite the zonal wind exhibiting a slight eastward vertical tilt. This eastward vertical tilt in the zonal wind field is consistent with the eastward tilt observed in the purely leaky $n = -1$ modes. This is because the barotropic mode leads the baroclinic mode, as can also be seen in the horizontal structure of the boundary layer vorticity decomposition in Figure \ref{f9}c.

Unfortunately the numerical model is quite unstable for the slow modes in the rigid-lid, surface friction limit. Despite strong sponge layers, instabilities unrelated to the slow modes (spurious noise, gravity waves) develop quickly in the numerical domain, precluding inferral of the complex growth rate and horizontal structure. The behavior of the barotropic mode in the slow modes will be discussed in the following section using the fully coupled, surface friction model, albeit with a large stratosphere stratification, instead of a rigid-lid.

These results show that surface friction act strictly as a damping mechanism in our framework. This is because the heating associated with frictional convergence is not divorced from the large-scale thermodynamics. Surface friction does, however, act to modify both the horizontal and vertical structure of the equatorial waves, as shown through the long poleward tails of the barotropic mode and vertical tilt in the zonal wind fields.


\subsection{Leaky, frictionally modified modes}
\begin{figure*}
  \includegraphics[width=39pc]{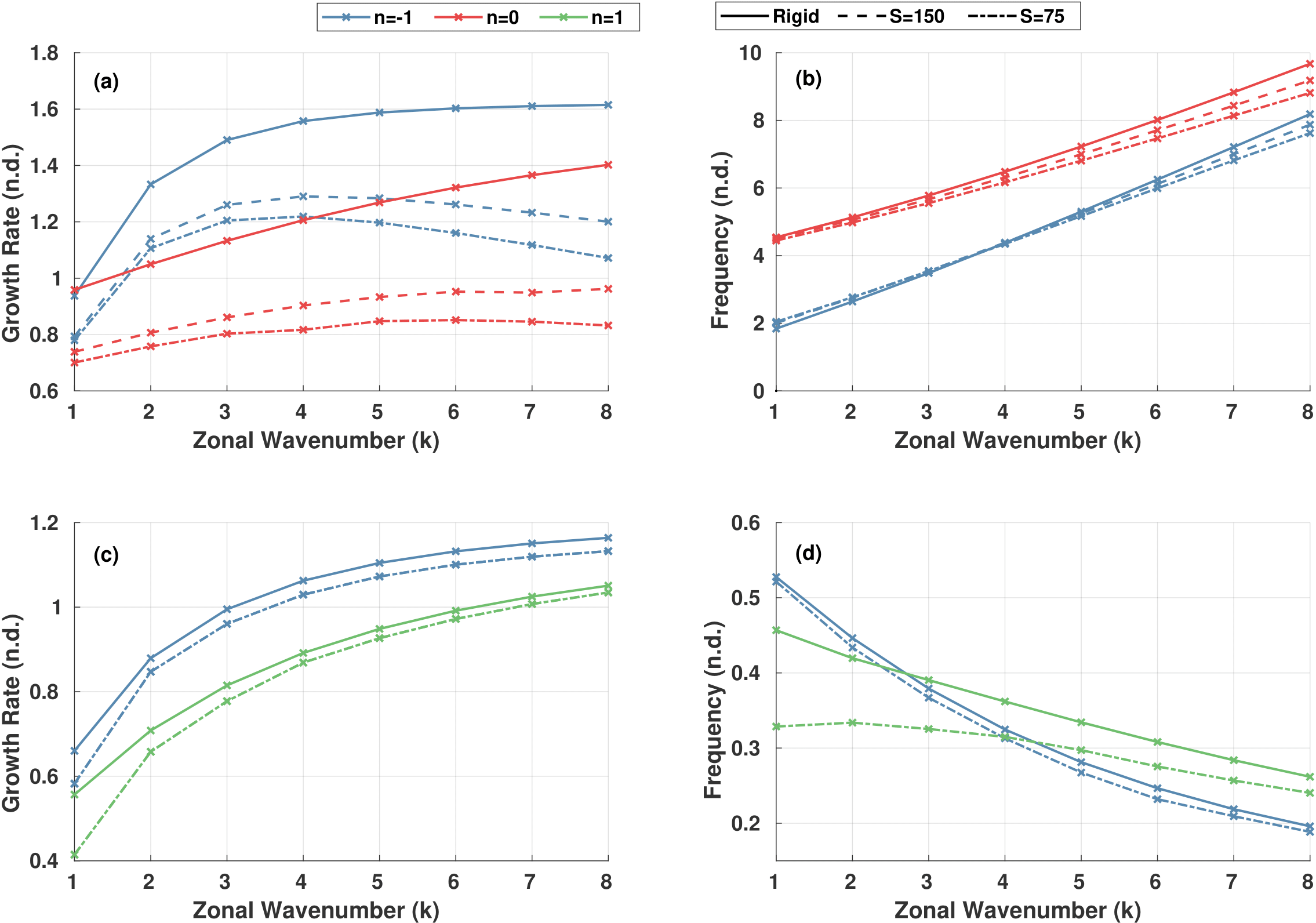}
  \caption{(a, c) Non-dimensional growth rate and (b, d) non-dimensional phase speed for the $n = -1, 0, 1$ modes of the (solid) rigid-lid system, the (dashed) surface friction + rigid-lid system with $S = 150$ and $F = 0.1$, and the (dot-dashed) surface friction + rigid-lid system with $S = 75$ and $F = 0.1$. Non-dimensional parameters selected for (a, b) are $\alpha = 3.5$, $\chi = 0.5$, $C = 0$, $\gamma = 1$, $D = 2.5$, $G = 0.25$, $\delta_x = 15$, and for (c, d) are $\alpha = 1$, $\chi = 1$, $C = 2.5$, $\gamma = 2$, $D = 1$, $G = 0.02$, $\delta_x = 30$. Only $S = 75$ case is shown for (c, d).} \label{f10}
\end{figure*}

While examining the excitation of the barotropic mode in the limits of surface friction under a rigid lid and a frictionless surface under a leaky lid were both useful exercises to isolate their respective effects on tropical waves, in the real world, both surface friction and leakage of energy to the stratosphere can act simultaneously to modify equatorial waves through excitation of the barotropic mode. In particular, the long-tails of the tropospherically trapped barotropic mode observed in the rigid-lid, surface friction model may behave differently if the barotropic mode can leak energy into the stratosphere.

To understand the extent to which both of these mechanisms can interact, we run the full troposphere-stratosphere numerical model with non-zero surface friction, using the same methods to infer the complex growth rates and eigenmodes. We first examine their combined effect on the WISHE-driven classical Matsuno modes through the same parameter set as used in Figures \ref{f5} and \ref{f8}. In addition, we focus on the eastward propagating modes since the westward propagating modes were not strongly affected by the presence of the stratosphere.

Figure \ref{f10}a and Figure \ref{f10}b show the non-dimensional growth rate and phase speeds for the WISHE destabilized Matsuno modes, respectively. Since there is both energy leakage into the stratosphere and surface friction, the growth rates are strongly dampened from the equivalent modes in the rigid-lid, inviscid limit. The frequencies of the WISHE Matsuno modes are not greatly modified, though all of the $n = 0$ modes and smaller scale $n = -1$ modes ($k \geq 5$) have slower phase speeds than their rigid-lid counterparts. Comparing with the growth rates in the stratosphere-only case shown in Figure \ref{f5}, we can see that the damping effects of the stratosphere and surface friction are approximately additive. 

\begin{figure*}
 \includegraphics[width=39pc]{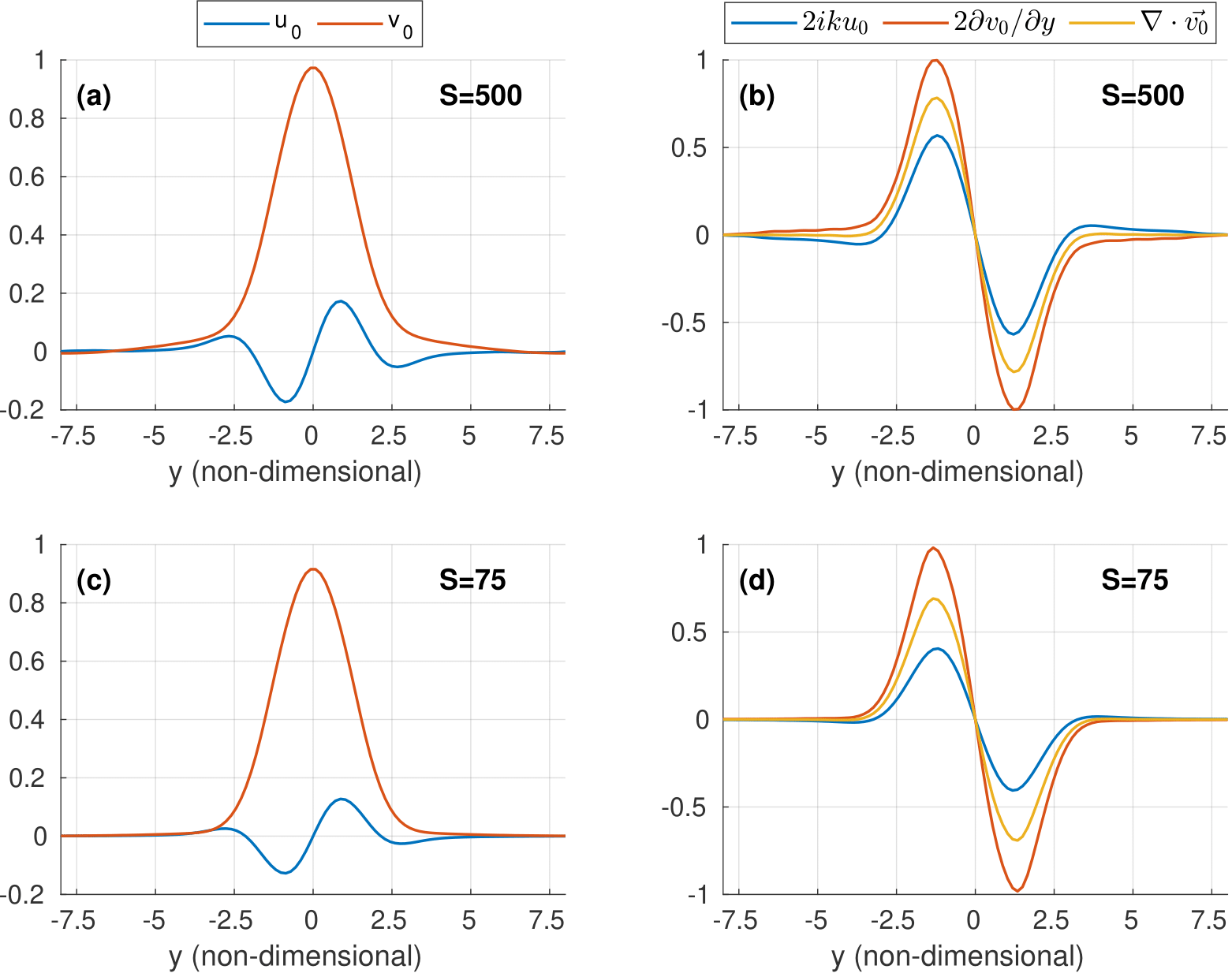}
  \caption{(a, c) Barotropic mode zonal and meridional velocities for the WISHE modified Matsuno $n = 0$, $k = 1$ mode using the same parameter set as in Figure \ref{f8}, $\alpha = 3.5$, $\chi = 0.5$, $C = 0$, $\gamma = 1$, $D = 2.5$, $G = 0.25$, $\delta_x = 15$, $F = 0.1$, but with (a) $S = 500$ and (c) $S = 75$. (b, d) $2 \pder[u_0]{x}$, $2 \pder[v_0]{y}$ (scaled by 2) and $\nabla \cdot \vec{v_0}$ for the same mode but with (b) $S = 500$ and (d) $S = 75$.} \label{f11}
\end{figure*}

The more interesting question, perhaps, is if the barotropic mode behaves differently when it can be excited by surface friction but also interact with the stratosphere. To examine this, we decompose the horizontal divergence of the barotropic mode into $\pder[u]{x}$, $\pder[v]{y}$, and the sum of both, which is equivalent to the pressure vertical velocity at the tropopause. Figure \ref{f11} shows the barotropic mode velocities and the barotropic horizontal divergence decomposition for the $n = 0$, $k = 1$ WISHE Matsuno mode, for a realistic stratosphere stratification ($S = 75$), as well as a highly stratified one ($S = 500$). In the case with a highly stratified stratosphere, we see the long-tail feature of the barotropic mode velocities, as also seen in the surface friction under a rigid-lid model. However, a key differing feature is that the barotropic mode velocities are no longer completely non-divergent, as evidenced in Figure \ref{f11}b. Further, the horizontal divergence of the barotropic mode (or, equivalently, the tropopause vertical velocity) decays to zero very quickly polewards, around $|y| = 3$. This is because near the equator, $\pder[u_0]{x}$ and $\pder[v_0]{y}$ have the same sign, evidence that the barotropic mode is exciting the stratosphere near the equator. In contrast, despite the barotropic velocities being small (though non-zero) polewards of the equator, the horizontal divergence is zero because $\pder[u_0]{x}$ and $\pder[v_0]{y}$ have opposite signs and almost exactly cancel. This is evidence of a tropospherically trapped barotropic mode. When we reduce the stratosphere stratification to $S = 75$, as shown in Figure \ref{f11}c and Figure \ref{f11}d, in effect allowing the tropopause to be more leaky, we see that the polewards extent of the barotropic mode velocities is greatly reduced, and hence the poleward extent of $\pder[u_0]{x}$ and $\pder[v_0]{y}$ is also reduced. The horizontal divergence still decays to zero around $|y| = 3$, as in the case where $S = 500$. Thus, when the troposphere is allowed to leak more energy into the stratosphere, the barotropic mode is no longer trapped in the troposphere, and the long-tail of the barotropic mode velocities is reduced. Finally, while we have taken a large stratification ($S = 500$) to emphasize the characteristics of the trapped barotropic mode, the extent of the trapping of the barotropic mode increases with $S$. Thus, in general, the polewards extent of the barotropic mode increases when the stratosphere stratification increases. 

Figure \ref{f10}c and Figure \ref{f10}d show the non-dimensional growth rate and phase speeds for the slow modes, for which we obtain by choosing the same parameter set as used in Figure \ref{f3} and \ref{f6}. Note, the $n = 0$ waves do not grow in time. The growth rates are not modified greatly from the rigid lid solutions; surface friction seems to have a weaker damping effect on the slower propagating modes. The damping effect of the stratosphere on the growth rates is small but the greatest at the largest scales, and almost negligible for the smaller scale waves, as before. However, the frequencies/phase speeds for the slow modes are reduced, by as much as 30\% for the $k = 1$ wave. The damping of the frequency is much greater for the $n = 1$ slow propagating modes than the $n = -1$ modes, and much stronger for larger scale $n = 1$ waves. It is also worth noting that the $S = 150$ case is not shown for the slow modes since the lines are nearly indistinguishable from the $S = 75$ case.

\begin{figure*}
  \noindent \includegraphics[width=39pc]{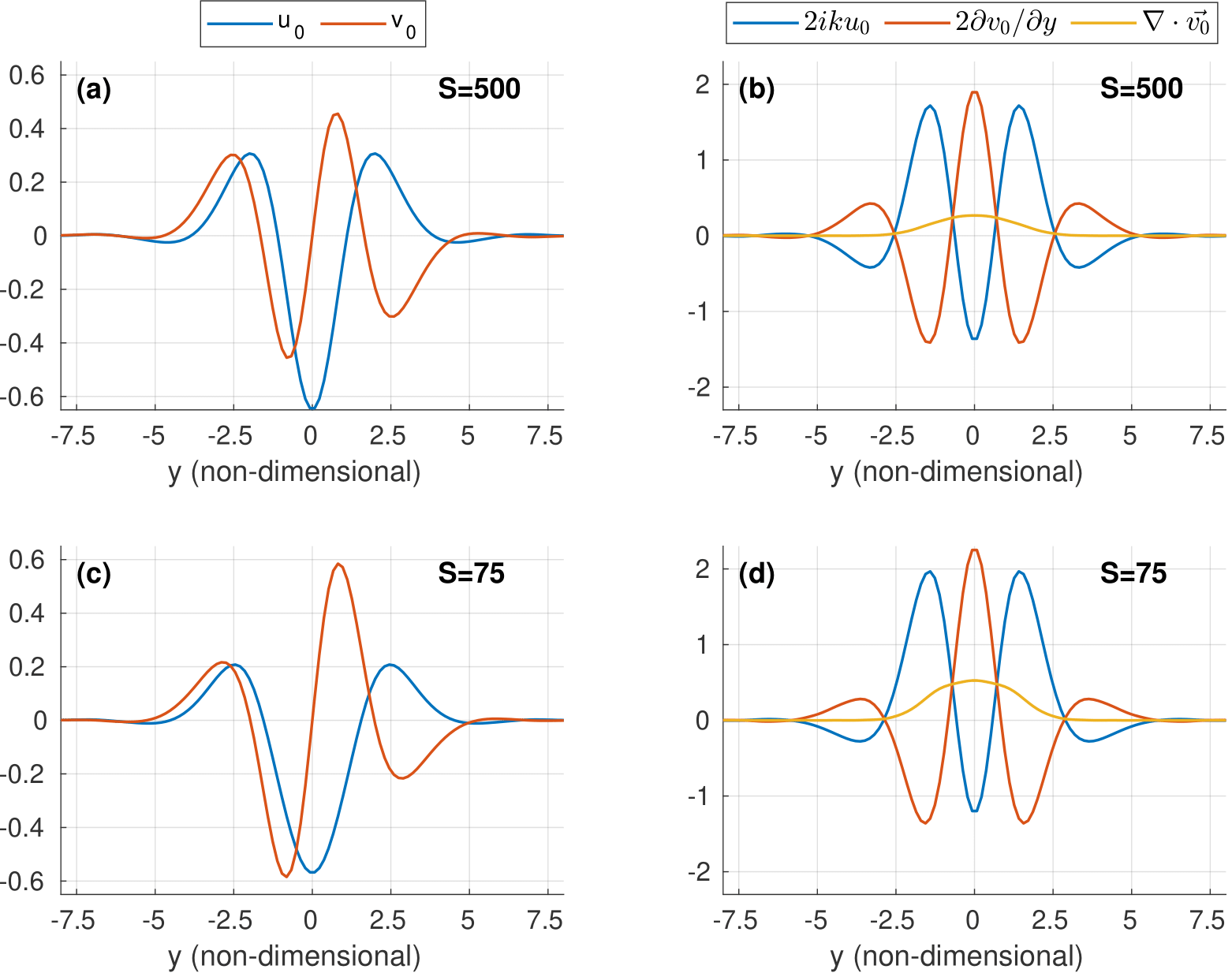}
  \caption{Analogous to Figure \ref{f11} but for the slow propagating $n = 1$, $k = 2$ mode using the parameter set $\alpha = 1$, $\chi = 1$, $C = 2.5$, $\gamma = 2$, $D = 1$, $G = 0.02$, $\delta_x = 30$, $F = 0.1$.} \label{f12}
\end{figure*}

The behavior of the poleward extent of the barotropic mode for the slow modes is similar to that for the WISHE-modified Matsuno modes. Figure \ref{f12} shows that the barotropic mode associated with the $n = 1$, $k = 2$ slow mode becomes non-divergent away from the equator and trapped in the troposphere. However, one key difference is that the slow modes do not leak much energy into the stratosphere to begin with, and thus decreasing the stratosphere stratification does not completely eliminate the long-tailed behavior of the barotropic velocities. This is evidenced by the small magnitude of the horizontal divergence of the barotropic mode (Figure \ref{f12}b, Figure \ref{f12}d), regardless of the stratosphere stratification, as $i k u_0$ and $\pder[v_0]{y}$ nearly cancel close to the equator ($|y| < 3$), and exactly cancel for $|y| > 3$.

What are the physical parameters that control the non-dimensional, $S$, and how can they vary across different equatorial waves? The non-dimensionalization of $S$ [see Appendix A] suggests that the stratosphere buoyancy frequency, $N^2$ and the meridional length scale, $L_y$, are quantities that could potentially lead to large variations in $S$. $L_y$ depends on a multitude of factors that can vary greatly in the tropics, such as the troposphere dry stratification, moist adiabatic lapse rate, and the precipitation efficiency. The largest influencing factor, however, is the precipitation efficiency, $\epsilon_p$: waves with larger $\epsilon_p$ experience a much greater stratosphere stratification. For instance, for Earth-like parameters and a $\epsilon_p = 0.5$, $S \approx 20$, while for $\epsilon_p = 0.95$, $S \approx 200$, an order of magnitude difference. While our definition of $\epsilon_p$ is an egregiously simple parameterization of cloud microphysics, the impact of $\epsilon_p$ on $S$ links cloud microphysical properties to the behavior of the barotropic mode.

\section{Discussion and summary \label{sec_summary}}
In this study, we extended a previously developed linear framework to include two mechanisms that can excite the barotropic mode in equatorial waves, surface friction, and coupling to the stratosphere. We first analyzed the modification of equatorial waves in the two separate limits of (1) coupling to the stratosphere with an inviscid surface, and (2) surface friction under a rigid-lid. Using a combination of theoretical solutions for the $v = 0$ mode and numerical solutions for higher order meridional modes, we found that the presence of a stratosphere leads to upward wave energy propagation that strongly dampens the growth rate of smaller-scale waves. This effect is consistent across the WISHE-modified Matsuno modes explored in this study, though it does not appreciably affect the slow propagating modes that are destabilized from cloud-radiative feedbacks. The barotropic mode is found to lead the baroclinic mode for the $n = -1$ eastward propagating modes, leading to a slight eastward tilt in the vertical that arises from the superposition of the two modes. The eastward tilt was found to be robust across horizontal scales and stratosphere stratification. In the limit of surface friction under a rigid-lid, we found that surface friction acts only to dampen growth rates by nearly a constant across all waves and zonal wavenumbers, which is reasonable given that surface friction acts on a constant time scale of damping in our framework. We also found that frictional excitation of the barotropic mode leads to long tails of non-divergent barotropic zonal and meridional velocities away from the equator. The barotropic mode is also found to lead the baroclinic mode for the $n = -1$ eastward propagating modes, leading to a vertical tilt in the zonal wind field, but not the vertical velocity field since the barotropic mode is non-divergent under a rigid lid.

The combined effects of surface friction and coupling to the stratosphere were analyzed using numerical solutions of the full linear model. We found that for the growth rates of the waves, the damping effects of surface friction and coupling to the stratosphere were approximately additive. Under a large non-dimensional stratosphere stratification, both the WISHE-modified Matsuno modes and slow modes exhibit tropospherically trapped barotropic modes away from the equator. Thus, the non-divergent barotropic velocities extend far away from the equator and are small but non-zero. When the non-dimensional stratosphere stratification is reduced, the poleward extent of the barotropic mode is greatly reduced. These results indicate that the dimensional variables that influence the non-dimensional stratosphere stratification, such as the buoyancy frequency in the stratosphere and precipitation efficiency, play key roles in modulating the poleward extent of the barotropic mode. While this study was restricted to theoretical analysis of the barotropic mode, future work will attempt to find evidence of the barotropic mode in both observational and numerical data.

This work models the interaction of equatorial waves with a zero mean flow stratosphere, and is the first basic step to illuminate how a dynamically dry and passive fluid influences the dynamics of a moist and convecting fluid underneath. Extension of the framework developed in this study to a non-zero mean flow in the stratosphere will be the subject of future work, which would allow for an investigation into the extent to which upwards wave radiation can explain the observed relationship between the MJO and Quasi-Biennial Oscillation (QBO) \citep{yoo2016modulation}.

Finally, it is worth discussing some of the short-comings of the modeling framework. Our application of surface friction in an infinitesimally small boundary layer with zero vertical velocity is exceptionally crude compared to real-world frictional boundary layers. In addition, the strict-quasi equilibrium approximation may not be as accurate for extremely short or high frequency waves \citep{ahmed2021quasi}, as evidenced by the presence of what looks like a second baroclinic mode in observations of convectively coupled Kelvin waves~\citep{straub2002observations}. Regardless, both surface friction and upward radiation of wave energy are shown to important mechanisms that influence the horizontal structure, vertical structure, and growth rates of equatorial waves, through excitation of the barotropic mode.

\acknowledgments
The authors gratefully acknowledge the support of the National Science Foundation through grant NSF ICER-1854929.

%
%
\datastatement
Model source code, instructions, and code to generate the figures from outputs of the numerical model are available at github.com/linjonathan. All code and figures were generated using Matlab.

%

\appendix[A]

\appendixtitle{Non-dimensionalization}
Here, we define the non-dimensional scalings for the variables that appear in the full linear model. The scalings for the tropospheric quantities $s^{*\prime}$, $s_m^\prime$, $\chi$, $\alpha$, $\gamma$, $D$ and $G$ are identical to those described in the appendix of KE18.
\begin{align}
    x &\rightarrow ax \\
    y &\rightarrow L_y y \\
    p &\rightarrow (p_s - p_t) p \\
    z^* &\rightarrow H z^* \\
    t &\rightarrow \frac{a}{\beta L_y^2} t \\
    L_y^4 &\rightarrow \frac{\Gamma_d}{\Gamma_m} \frac{d \overline{s}_d}{dz} (T_b - [\overline{T}]) H \frac{1 - \epsilon_p}{\beta^2} \\
    u^\prime &\rightarrow \frac{a C_k |\overline{\textbf{V}}|}{H}  u^\prime \\
    v^\prime &\rightarrow \frac{L_y C_k |\overline{\textbf{V}}|}{H}  v^\prime \\
    w^{*\prime} &\rightarrow  C_k |\overline{\textbf{V}}|  w^{*\prime} \\
    \phi^\prime &\rightarrow \frac{a \beta L_y^2 C_k |\overline{\textbf{V}}| }{H} \phi^\prime \\
    \omega &\rightarrow \frac{C_k \overline{\textbf{V}} (p_s - p_t)}{H} \omega \\
    N^2 &\rightarrow \frac{\beta^2 L_y^4}{H^2} S  \\
    F &\rightarrow \frac{a C_d |\overline{\textbf{V}}|}{\beta L_y^2 h_b}
\end{align}
where most dimensional parameters are described in the main text and in KE18. Parameters not defined in this text are the mean radius of the Earth $a$, the dry adiabatic lapse rate $\gamma_d$, the moist adiabatic lapse rate $\gamma_m$, the dry entropy stratification $\frac{d \overline{s}_d}{dz}$, the precipitation efficiency $\epsilon_p$, and the enthalpy exchange coefficient $C_k$. The terms on the left of the arrow are the dimensional quantities, and those on the right are the non-dimensional quantities.

\appendix[B]

\appendixtitle{Numerical Model}
The full mathematical description of the numerical system (including damping terms) is below:
\begin{align}
    \pder[u_0]{t} &= -ik \big[ \phi_{s} + V_1(p_t) s^* \big] + y v_0 - 2 F (u_0 + u_1) - r u_0 \\ 
    \pder[v_0]{t} &= \delta_x \Big[ -\pder[]{y} \big[ \phi_{s} + V_1(p_t) s^* \big] - y u_0 \Big] - F (v_0 + v_1) - r v_0 \\ 
    \pder[u_1]{t} &= i k s^* + y v_1 - 2 F (u_0 + u_1) - r u_1 \\ 
    \pder[v_1]{t} &= \delta_x \Big[ \pder[s^*]{y} - y u_1 \Big] - F (v_0 + v_1)  - r v_1 \\ 
    \pder[s^*]{t} &= (1 + C) s_m - w - \alpha (u_0 + u_1) - \chi s^* - r s^* \\ 
    \gamma \pder[s_m]{t} &= -D s^* - \alpha (u_0 + u_1) - G w + C s_m  - r s_m \\ 
    w &= -ik(u_0 + u_1) - \pder[]{y} (v_0 + v_1)\\
    \pder[u_s]{t} &= -ik \phi_s + y v_s - r u_s  \\ 
    \pder[v_s]{t} &= \delta_x \Big[ -\pder[\phi_s]{y} - y u_s \Big] - r v_s \\ 
    \pder[\phi_s]{t} &= - \int^{z}_\infty w^*_s S \: dz^* - r \phi_s \\ 
    \rho_s w^*_s &= -B \Big( ik u_0 + \pder[v_0]{y} \Big) - \int_{z^* = 1}^{z} \Big[ \rho_s \Big( i k u_s(y, z^*) + \pder[]{y} v_s(y, z^*) \Big) \Big] dz^*
\end{align}
where all variables are defined in the main text with the exception of $r$, which is the sponge coefficient for the sponge layer that is applied at the boundaries of the domain.




%
%
%
\bibliographystyle{ametsoc2014}
\bibliography{references}

%

%

\end{document}